\documentclass[11pt,letterpaper,twocolumn]{article}
\usepackage[utf8]{inputenc}
\usepackage[T1]{fontenc}
\usepackage{amsmath}
\usepackage{amsfonts}
\usepackage{amssymb}
\usepackage{fullpage}
\usepackage{bbold}
\usepackage{xcolor}
\usepackage{natbib}
\usepackage[font={small}]{caption}
\usepackage[affil-it]{authblk}

\usepackage{longtable}
\usepackage{float}
\usepackage[hmargin={0.6in,0.6in},vmargin={1in,1in}]{geometry}
\usepackage[space]{grffile}
\usepackage{subcaption}
\usepackage{graphicx}
\usepackage{epstopdf} 
\usepackage{algpseudocode}
\usepackage{booktabs}

\author[1,2,3]{J. Doyne Farmer}
\author[1,4,5]{Fran\c{c}ois Lafond}
\affil[1]{Institute for New Economic Thinking at the Oxford Martin School, University of Oxford, Oxford OX2 6ED, U.K. }
\affil[2]{Mathematical Institute, University of Oxford, Oxford OX1 3LP, U. K.}
\affil[3]{Santa-Fe Institute, Santa Fe, NM 87501, U.S.A}
\affil[4]{London Institute for Mathematical Sciences, London W1K 2XF, U.K.}
\affil[5]{United Nations University - MERIT, 6211TC Maastricht, The Netherlands}

\title{How predictable is technological progress?
\footnote{Acknowledgements: We would like to acknowledge Diana Greenwald and Aimee Bailey for their help in gathering and selecting data, as well as Glen Otero for help acquiring data on genomics, Chris Goodall for bringing us up to date on developments in solar PV, and Christopher Llewellyn Smith, Jeff Alstott, Michael Totten, our INET colleagues and three referees for comments.  This project was supported by the European Commission project {FP7-ICT-2013-611272} (GROWTHCOM) and by the U.S. Dept. of Solar Energy Technologies Office under grant {DE-EE0006133}. Contacts: doyne.farmer@inet.ox.ac.uk; francois.lafond@inet.ox.ac.uk  }
}

\begin{document}

\maketitle

\begin{abstract}
Recently it has become clear that many technologies follow a generalized version of Moore's law, i.e. costs tend to drop exponentially, at different rates that depend on the technology.  Here we formulate Moore's law as a correlated geometric random walk with drift, and apply it to historical data on 53 technologies.  We derive a closed form expression approximating the distribution of forecast errors as a function of time.  Based on hind-casting experiments we show that this works well, making it possible to collapse the forecast errors for many different technologies at different time horizons onto the same universal distribution.  This is valuable because it allows us to make forecasts for any given technology with a clear understanding of the quality of the forecasts.  As a practical demonstration we make distributional forecasts at different time horizons for solar photovoltaic modules, and show how our method can be used to estimate the probability that a given technology will outperform another technology at a given point in the future. 



\end{abstract}


\section{Introduction}
\label{section:intro}

Technological progress is widely acknowledged as the main driver of economic growth, and thus any method for improved technological forecasting is potentially very useful. Given that technological progress depends on innovation, which is generally thought of as something new and unanticipated, forecasting it might seem to be an oxymoron.  In fact there are several postulated laws for technological improvement, such as Moore's law and Wright's law, that have been used to make predictions about technology cost and performance.  But how well do these methods work?

\begin{figure}[H]
\includegraphics[height=74mm]{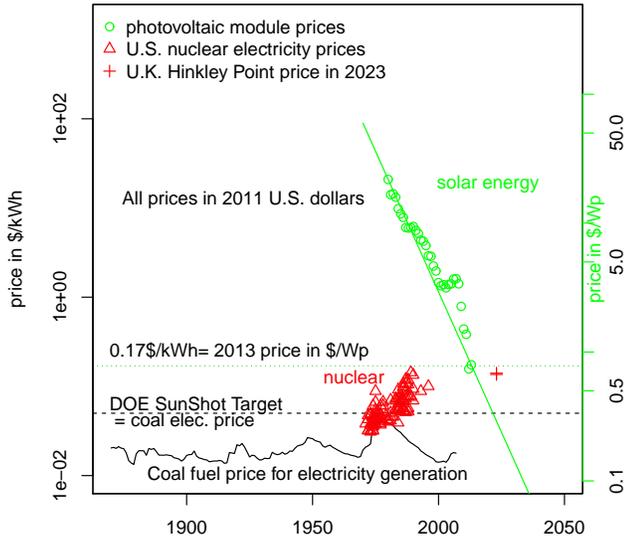}
\caption{{\it A comparison of long-term price trends for coal, nuclear power and solar photovoltaic modules}.  Prices for coal and nuclear power are costs in the US in dollars per kilowatt hour (scale on the left) whereas solar modules are in dollars per watt-peak, i.e. the cost for the capacity to generate a watt of electricity in full sunlight (scale on the right). For coal we use units of the cost of the coal that would need to be burned in a modern US plant if it were necessary to buy the coal at its inflation-adjusted price at different points in the past.  Nuclear prices are Busbar costs for US nuclear plants in the year in which they became operational (from \citet{Cooper}).  The alignment of the left and right vertical axes is purely suggestive; based on recent estimates of levelized costs, we took $\$0.177/\mbox{kWh} = \$0.82/\mbox{Wp}$ in 2013 (2013\$). The number \$0.177/\mbox{kWh} is a global value produced as a projection for 2013 by the International Energy Agency (Table 4 in \citet{IEA2014}). We note that it is compatible with estimated values (Table 1 in \citet{baker2013economics}, Fig. 4 in \citet{IEA2014}). The red cross is the agreed price for the planned UK Nuclear power plant at Hinkley Point which is scheduled to come online in 2023 (\pounds0.0925 $\approx$ \$0.14). The dashed line corresponds to an earlier target of $\$0.05/\mbox{kWh}$ set by the the U.S. Department of Energy.}
\label{fig:electricity}
\end{figure}

Predictions are useful because they allow us to plan, but to form good plans it is necessary to know probabilities of possible outcomes.  Point forecasts are of limited value unless they are very accurate, and when uncertainties are large they can even be dangerous if they are taken too seriously.   At the very least one needs error bars, or better yet, a distributional forecast, estimating the likelihood of different future outcomes.  Although there are now a few papers testing technological forecasts\footnote{
See e.g. \citet{alchian1963reliability}, \citet{alberth2008forecasting}.  \citet{nagy2013statistical} test the relative accuracy of different methods of forecasting statistically but do not produce and test a distributional estimate of forecast reliability for any particular method. McCrory, cited in \citet{jantsch1967technological}, assumes a Gaussian distribution and uses this to calculate the probability that a targeted level of progress be met at a given horizon. Here we assume and test a Gaussian distribution for the natural log.}
there is as yet no method that gives distributional forecasts based on an empirically validated stochastic process.  In this paper we remedy this situation by deriving the distributional errors for a simple forecasting method and testing our predictions on empirical data on technology costs.
To motivate the problem that we address, consider three technologies related to electricity generation:  coal mining, nuclear power and photovoltaic modules.  Fig.~\ref{fig:electricity} compares their long-term historical prices.   Over the last 150 years the inflation-adjusted price of coal has fluctuated by a factor of three or so, but shows no long term trend, and indeed from the historical time series one cannot reject the null hypothesis of a random walk with no drift\footnote{
	To drive home the point that fossil fuels show no long term trend of dropping in cost, after adjusting for inflation coal now costs about what it did in 1890, and a similar statement applies to oil and gas.}
\citep{McNerney2011historical}.   Nuclear power and solar photovoltaic electricity, in contrast, are both new technologies that emerged at roughly the same time.  The first commercial nuclear power plant opened in 1956 and the first practical use of solar photovoltaics was as a power supply for the Vanguard I satellite in 1958.   The cost of electricity generated by nuclear power is highly variable, but has generally increased by a factor of two or three during the period shown here.  In contrast, a watt of solar photovoltaic capacity cost \$256 in 1956 \citep{perlin1999space} (about \$1910 in 2013 dollars) vs. \$0.82 in 2013, dropping in price by a factor of about 2,330. Since 1980 photovoltaic modules have decreased in cost at an average rate of about $10\%$ per year.

In giving this example we are not trying to make a head-to-head comparison of the full system costs for generating electricity.  Instead we are comparing three different technologies, coal mining, nuclear power and photovoltaic manufacture.  Generating electricity with coal requires plant construction (whose historical cost has dropped considerably since the first plants came online at the beginning of the 20th century).  Generating electricity via solar photovoltaics has balance of system costs that have not dropped as fast as that of modules in recent years.  Our point here is that different technologies can decrease in cost at very different rates.

Predicting the rate of technological improvement is obviously very useful for planning and investment.  But how consistent are such trends?  In response to a forecast that the trends above will continue, a skeptic would rightfully respond, ``How do we know that the historical trend will continue?  Isn't it possible that things will reverse, and over the next 20 years coal will drop in price dramatically and solar will go back up?".

Our paper provides a quantitative answer to this question.  We put ourselves in the past, pretend we don't know the future, and use a simple method to forecast the costs of 53 different technologies.  Actually going through the exercise of making out-of-sample forecasts rather than simply doing in-sample regressions has the essential advantage that it fully mimics the process of making forecasts and allows us to say precisely how well forecasts would have performed.  Out-of-sample testing such as we do here is particularly important when models are mis-specified, which one expects for a complicated phenomenon such as technological improvement.

We show how one can combine the experience from forecasting many technologies to make reliable distributional forecasts for a given technology.  For solar PV modules, for example, we can say, ``Based on experience with many other technologies, the probability is roughly $5\%$ that in 2030 the price of solar PV modules will be greater than or equal to their current (2013) price".   We can assign a probability to different price levels at different points in the future, as is done later in Fig.~\ref{fig:solarforecast} (where we show that very likely the price will drop significantly).   We can also compare different technologies to assess the likelihood of different future scenarios for their relative prices, as is done in Fig.~\ref{fig:psn}.

Technological costs occasionally experience structural breaks where trends change.  Indeed there are several clear examples in our historical data, and although we have not explicitly modeled this, their effect on forecast errors is included in the empirical analysis we have done here.  The point is that, while such structural breaks happen, they are not so large and so common as to over-ride our ability to forecast.  Every technology has its own story, its own specific set of causes and effects, that explain why costs went up or down in any given year.  Nonetheless, as we demonstrate here, the long term trends tend to be consistent, and can be captured via historical time series methods with no direct information about the underlying technology-specific stories.

In this paper we use a very simple approach to forecasting that was originally motivated by Moore's Law.  As everyone knows, Intel's ex-CEO, Gordon Moore, famously predicted that the number of transistors on integrated circuits would double every two years, i.e. at an annual rate of about 40\%.   Making transistors smaller also brings along a variety of other benefits, such as increased speed, decreased power consumption, and less expensive manufacture costs per unit of computation.   As a result it quickly became clear that Moore's law applies more broadly, for example, implying a doubling of computational speed every 18 months.  

Moore's law stimulated others to look at related data more carefully, and they discovered that exponential improvement is a reasonable approximation for other types of computer hardware as well, such as hard drives.  Since the performance of hard drives depends on physical factors that are unrelated to transistor density this is an independent fact, though of course the fact that mass storage is essential for computation causes a tight coupling between the two technologies.  Lienhard, Koh and Magee, and others\footnote{
Examples include \citet{lienhard2006invention}, \citet{koh2006functional,koh2008functional,bailey2012forecasting,benson2014improvement,benson2014quantitative}, \citet{nagy2013statistical}.  Studies of improvement in computers over long spans of time indicate super-exponential improvement \citep{nordhaus2007two, Nagy2}, suggesting that Moore's law may only be an approximation reasonably valid over spans of time of 50 years or less. See also e.g. \citet{funk2013drives} for an explanation of Moore's law based on geometric scaling, and \citet{funk2014rapid} for empirical evidence regarding fast improvement prior to large production increase.}
examined data for other products, including many that have nothing to do with computation or information processing, and postulated that exponential improvement is a much more general phenomenon that applies to many different technologies, even if in most cases the exponential rates are much slower.

Although Moore's law is traditionally applied as a regression of the log of the cost on a deterministic time trend, we reformulate it here as a geometric random walk with drift.  This has several advantages. On average it results in more accurate forecasts, especially at short horizons, indicating that it is indeed a better model. In addition, this allows us to use standard results from the time series forecasting literature\footnote{
Several methods have been defined to obtain prediction intervals, i.e. error bars for the forecasts \citep{chatfield1993calculating}. The classical Box-Jenkins methodology for ARIMA processes uses a theoretical formula for the variance of the process, but does not account for uncertainty due to parameter estimates. Another approach is to use the empirical forecast errors to estimate the distribution of forecast errors. In this case, one can use either the in-sample errors (the residuals, as in e.g. \citet{taylor1999quantile}), or the out-of-sample forecast errors \citep{williams1971simple,lee2014empirical}. Several studies have found that using residuals leads to prediction intervals which are too tight \citep{makridakis1989sampling}.}.
The technology time series in our sample are typically rather short, often only 15 or 20 points long, so to test hypotheses it is essential to pool the data.   Because the geometric random walk is so simple it is possible to derive formulas for the forecast errors in closed form.  This makes it possible to estimate the forecast errors as a function of both sample size and forecasting horizon, and to combine data from many different technologies into a single analysis.  This allows us to get highly statistically significant results.  And most importantly, because this allows us to systematically test the method on data for many different technologies, this allows us to make distributional forecasts for a single technology and have confidence in the results.

Motivated by structure we find in the data, we further extend Moore's law to allow for the possibility that changes in price are positively autocorrelated in time.  We assume that the logarithm of the cost follows a random walk with drift and autocorrelated noise, more specifically an Integrated Moving Average process of order (1,1), i.e. an IMA(1,1) model. Under the assumption of sufficiently large autocorrelation this method produces a good fit to the empirically observed forecasting errors.  We derive a formula for the errors of this more general model, assuming that all technologies have the same autocorrelation parameter and the forecasts are made using the simple random walk model.
We use this to forecast the likely distribution of the price of photovoltaic solar modules, and to estimate the probability that solar modules will undercut a competing technology at a given date in the future.

We want to stress that we do not mean to claim that the generalizations of Moore's law explored here provide the most accurate possible forecasts for technological progress.  There is a large literature on experience curves\footnote{
 \citet{arrow1962economic,alchian1963reliability,argote1990learning,dutton1984treating,thompson2012relationship}.},
studying the relationship between cost and cumulative production originally suggested by Wright (\citeyear{wright1936factors}), and many authors have proposed alternatives and generalizations\footnote{
See \citet{goddard1982debunking,sinclair2000what,jamasb2007technical,nordhaus2009perils}. }.
\citet{nagy2013statistical} tested these alternatives using a data set that is very close to ours and found that Moore's and Wright's laws were roughly tied for first place in terms of their forecasting performance.  An important caveat is that Nagy et al.'s study was based on a trend stationary model, and as we argue here, the difference stationary model is superior, both for forecasting and for statistical testing.  It seems likely that methods using auxiliary data such as production, patent activity, or R\&D can be used to make forecasts for technological progress that incorporate more factors, and that such methods should yield improvements over the simple method we use here\footnote{
See for example \cite{benson2014quantitative} for an example of how patent data can be used to explain variation in rates of improvement among different technologies.
}.

The key assumption made here is that all technologies follow the same random process, even if the drift and volatility parameters of the random process are technology specific. This allows us to develop distributional forecasts in a highly parsimonious manner and efficiently test them out of sample.  We restrict ourselves to forecasting unit cost in this paper, for the simple reason that we have data for it and it is comparable across different technologies.  The work presented here provides a simple benchmark against which to compare forecasts of future technological performance based on other methods.  

The approach of basing technological forecasts on historical data that we pursue here stands in sharp contrast to the most widely used method, which is based on expert opinions.  The use of expert opinions is clearly valuable, and we do not suggest that it should be supplanted, but it has several serious drawbacks.  Expert opinions are subjective and can be biased for a variety of reasons \citep{albright2002can}, including common information, herding, or vested interest.  Forecasts for the costs of nuclear power in the US, for example, were for several decades consistently low by roughly a factor of three \citep{Cooper}.  A second problem is that it is very hard to assess the accuracy of expert forecasts.   In contrast the method we develop here is objective and the quality of the forecasts is known.  Nonetheless we believe that both methods are valuable and that they should be used side-by-side.\footnote{For additional discussion of the advantages and drawbacks of different methods of technology forecasting, see \citet{ayres1969technological}, \citet{martino1993technological} and  \citet{NRC}}

The remainder of the paper develops as follows:  In Section~\ref{models} we derive the error distribution for forecasts based on the geometric random walk as a function of time horizon and other parameters and show how the data for different technologies and time horizons should be collapsed.  We also show how this can be generalized to allow for autocorrelations in the data and derive similar (approximate) formulas.  In Section~\ref{data} we describe our data set and present an empirical relationship between the variance of the noise and the improvement rate for different technologies. In Section~\ref{section:estim} we describe our method of testing the models against the data, and present the results in Section~\ref{empiricalResults}.  We then apply our method to give a distributional forecast for solar module prices in Section~\ref{comparison} and show how this can be used to forecast the likelihood that one technology will overtake another.  Finally we give some concluding remarks in Section~\ref{conclusion}.  A variety of technical results are given in the appendices.

\section{Models\label{models}}

\subsection{Geometric random walk}

In this section we discuss how to formulate Moore's law in the presence of noise and argue that the best method is the geometric random walk with drift.  We then present a formula for the distribution of expected errors as a function of the time horizon and the other parameters of the model, and generalize the formula to allow for autocorrelation in the data generating process.  This allows us to pool the errors for many different technologies.  This is extremely useful because it makes it possible to test the validity of these results using many short time series (such as the data we have here).

The generalized version of Moore's law we study here is a postulated relationship which in its deterministic form is
\[
p_t=p_0e^{\mu t},
\]
where $p_t$ is either the unit cost or the unit price of a technology at time $t$; we will hereafter refer to it as the {\it cost}.   $p_0$ is the initial cost and $\mu$ is the exponential rate of change.   (If the technology is improving then $\mu < 0$.)  In order to fit this to data one has to allow for the possibility of errors and make an assumption about the structure of the errors.  Typically the literature has treated Moore's law using a trend stationary model, minimizing squared errors to fit a model of the form
\begin{equation}
y_t=y_0+\mu t+e_t,
\label{eq:yt1}
\end{equation}
where $y_t=\log(p_t)$.  From the point of view of the regression, $y_0$ is the intercept, $\mu$ is the slope and $e_t$ is independent and identically distributed (IID) noise.  

But records of technological performance such as those we study here are time series, giving the costs $p_{jt}$ for technology $j$ at successive times $t = 1, 2, \ldots, T_j$.  It is therefore more natural to use a time series model.  The simplest possible choice that yields Moore's law in the deterministic limit is the geometric random walk with drift,
\begin{equation}
y_{t} = y_{t-1} + \mu +n_t.
\label{eq:yt2}
\end{equation}
As before $\mu$ is the drift and $n_t$ is an IID noise process.  Letting the noise go to zero recovers the deterministic version of Moore's law in either case.  When the noise is nonzero, however, the models behave quite differently. For the trend stationary model the shocks are purely transitory, i.e. they do not accumulate.  
In contrast, if $y_0$ is the cost at time $t = 0$, Eq. (\ref{eq:yt2}) can be iterated and written in the form 
\begin{equation}
y_t=y_0+\mu t+\sum_{i=1}^{t}n_i.
\label{eq:ytasSum}
\end{equation}
This is equivalent to  Eq. (\ref{eq:yt1}) except for the last term. 
While in the regression model of Eq. (\ref{eq:yt1}) the value of $y_t$ depends only on the current noise and the slope $\mu$, in the random walk model (Eq. \ref{eq:yt2}) it depends on the sum of previous shocks. Hence shocks in the random walk model accumulate and the forecasting errors grow with time horizon as one would expect, even if the parameters of the model are perfectly estimated.\footnote{
\citet{nagy2013statistical} used trend stationary models to study a similar dataset. Their short term forecasts were on average less accurate and they had to make ad hoc assumptions to pool data from different horizons.
}.

For time series models a key question is whether the process has a unit root. Most of our time series are much too short for unit root tests to be effective \citep{blough1992relationship}.
Nonetheless, we found that our time series forecasts are consistent with the hypothesis of a unit root and that they perform better than several alternatives.

\subsection{Prediction of forecast errors}

We now derive a formula for the forecast errors of the geometric random walk as a function of time horizon.  We assume that all technologies follow the geometric random walk, i.e. our noisy version of Moore's law, but with technology-specific parameters.   Rewriting Eq. (\ref{eq:yt2}) slightly, it becomes 
\[
y_{jt} = y_{j,(t-1)} + \mu_j +n_{jt},
\]
where the index $j$ indicates technology $j$.  For convenience we assume that noise $n_{jt}$ is IID normal, i.e. $n_{jt} \sim \mathcal{N}(0,K_j^2)$.  This means that technology $j$ is characterized by a drift $\mu_j$ and the standard deviation of the noise increments $K_j$.   We will typically not include the indices for the technology unless we want to emphasize the dependence on the technology.

We now derive the expected error distribution for Eq. (\ref{eq:yt2}) as a function of the time horizon $\tau$.
Eq. (\ref{eq:yt2}) implies that
\begin{equation}
y_{t+\tau}=y_t+ \mu \tau +  \sum_{i=t+1}^{t+\tau}n_i.
\label{eq:MRWttotau}
\end{equation}
The point forecast $\tau$ steps ahead is\footnote{
The point forecast is the expected logarithm of the cost for the random walk with drift model, $E[y_{t + \tau}]$.  We assume $y_{t + \tau}$ is normally distributed. This means the cost is log-normally distributed and the forecast of the median cost is $e^{E[y_{t + \tau}]}$.  Because the mean of a log-normal distribution also depends on the variance of the underlying normal distribution, the expected cost diverges when $\tau \to \infty$ due to parameter uncertainty.  Our forecasts here are for the median cost.  This has the important advantage that (unlike the mean or the mode) it does not require an estimate of the variance, and is therefore simpler and more robust.
%
}
\begin{equation}
\hat{y}_{t+\tau}=y_t +\hat{\mu} \tau,
\label{eq:predictionMRW}
\end{equation}
where $\hat{\mu}$ is the estimated $\mu$. The forecast error is defined as 
\begin{equation}
\mathcal{E}=y_{t+\tau}-\hat{y}_{t+\tau}.
\label{eq:error2}
\end{equation}
Putting Eqs. (\ref{eq:MRWttotau}) and (\ref{eq:predictionMRW}) into Eq.~(\ref{eq:error2}) gives
\begin{equation}
\mathcal{E}=\tau(\mu-\hat{\mu})+\sum_{i=t+1}^{t+\tau}n_i,
\label{eq:error}
\end{equation}
which separates the error into two parts. The first term is the error due to the fact that the mean is an estimated parameter and the second term represents the error due to the fact that unpredictable random shocks accumulate \citep{sampson1991effect}.  Assuming that the noise increments are i.i.d normal and that the estimation of the parameters is based on a trailing sample of $m$ data points, in Appendix~\ref{appendix:forecasterrorsRW} we derive the scaling of the errors with $m$, $\tau$ and $\hat{K}$, where $\hat{K}^2$ is the \emph{estimated} variance.

Because we want to aggregate forecast errors for technologies with different volatilities, to study how the errors grow as a function of $\tau$ we use the \emph{normalized} mean squared forecast error $\Xi (\tau)$.  Assuming $m>3$  it is
\begin{equation}
\Xi (\tau) \equiv E \left [ \left(\frac{\mathcal{E}}{\hat{K}} \right)^2\right]=\frac{m-1}{m-3} \hspace{1mm} \left(\tau+\frac{\tau^2}{m}\right), 
\label{MSFEmain}
\end{equation}
where $E$ represents the expectation.

This formula makes intuitive sense.  The diffusion term $\tau$  is due to the accumulation of noisy fluctuations through time.  This term is present even in the limit $m \to \infty$, where the estimation is perfect.  The $\tau^2/m$ term is due to estimation error in the mean.  The need to estimate the variance causes the prefactor\footnote{
The prefactor is significantly different from one only when $m$ is small. \cite{sampson1991effect} derived the same formula but without the prefactor since he worked with the true variance. \cite{sampson1991effect} also showed that the square term due to error in the estimation of the drift exists for the regression on a time trend model, and for more general noise processes. See also \citet{clements2001forecasting}.}
$(m - 1)/(m-3)$ 
and also means that the distribution is Student $t$ rather than normal, i.e.
\begin{equation}
\epsilon=\frac{1}{\sqrt{A}} \left(\frac{\mathcal{E}}{\hat{K}}\right) \sim t(m-1),
\label{eq:distrt}
\end{equation}
with
\begin{equation}
A=\tau + \tau^2/m.
\label{eq:A}
\end{equation}
Eq.~(\ref{eq:distrt}) is universal in the sense that the right hand side is independent of $\hat{\mu}_j$, $\hat{K}_j$, and $\tau$.  It depends neither on the properties of the technology nor on the time horizon.  As a result we can pool forecast errors for different technologies at different time horizons.  This property is extremely useful for statistical testing and can also be used to construct distributional forecasts for a given technology.

\subsection{Generalization for autocorrelation\label{SecIMA}}

We now generalize the formula above to allow for autocorrelations in the error terms.  Although the uncorrelated random walk model above does surprisingly well, there is good evidence that there are positive autocorrelations in the data.  In order to incorporate this structure we extend the results above for an ARIMA(0,1,1)  (autoregressive integrated moving average) model.  The zero indicates that we do not use the autoregressive part, so we will abbreviate this as an IMA(1,1) model in what follows.  The IMA(1,1) model is of the form
\begin{equation}
y_{t}-y_{t-1}=\mu+v_t+\theta v_{t-1},
\label{eq:IMA11}
\end{equation}
with the noise $v_t\sim \mathcal{N}(0,\sigma^2)$.  This model is also a geometric random walk, but with correlated increments when $\theta \ne 0$ (the autocorrelations of the time series are positive when $\theta > 0$).

We chose this model rather than other alternatives mainly for its simplicity\footnote{
Our individual time series are very short, which makes it very difficult to find the proper order of differencing and to distinguish between different ARMA models. For instance, slightly different ARIMA models such as (1,1,0) are far from implausible for many technologies.}.
Moreover, our data are often time-aggregated, that is, our yearly observations are averages of the observed costs over the year. It has been shown that if the true process
is a random walk with drift then aggregation can lead to substantial autocorrelation \citep{working1960note}. In any case, while every technology certainly follows an idiosyncratic pattern and may have a complex autocorrelation structure and specific measurement errors, using the IMA(1,1) as a universal model allows us to parsimoniously understand the empirical forecast errors and generate robust prediction intervals.

A key quantity for pooling the data is the variance, which by analogy with the previous model we call $K$ for this model as well.  It is easy to show that
\[
K^2 \equiv var(y_{t}-y_{t-1})=var(v_t+\theta v_{t-1})=(1+\theta^2)\sigma^2,
\] 
see e.g. \cite{box1970time}.  
The relevant formulas for this case are derived in Appendix~\ref{appendix:forecasterrorsIMA}.  We make the same point forecasts as before given by Eq.~(\ref{eq:predictionMRW}).  If the variance is known the distribution of forecast errors is 
\begin{equation}
\mathcal{E} \sim\mathcal{N}(0,\sigma^2 A^*),
\label{eq:eIMAnorm}
\end{equation}
with
\begin{equation}
A^*=-2 \theta + \left(1+\frac{2 (m-1) \theta }{m}+\theta ^2\right) \left(\tau + \frac{\tau^2}{m}\right).
\label{eq:Astar}
\end{equation}
Note that we recover Eq. (\ref{eq:A}) when $\theta=0$. 
In the usual case where the variance has to be estimated, we derive an approximate formula for the growth and distribution of the forecast errors by assuming that $\hat{K}$ and $\mathcal{E}$ are independent. The expected mean squared normalized  error is
\begin{equation}
\Xi (\tau) \equiv E \left [ \left(\frac{\mathcal{E}}{\hat{K}} \right)^2\right]=\frac{m-1}{m-3} \hspace{1mm}\frac{A^*}{1+\theta^2},
\label{MSFEmainIMA}
\end{equation}
and the distribution of rescaled normalized forecast errors is
\begin{equation}
\epsilon^*=\frac{1}{\sqrt{A^*/(1+\theta^2)}} \left(\frac{\mathcal{E}}{\hat{K}}\right) \sim t(m-1).
\label{eq:distrtMA}
\end{equation}

These formulas are only approximations so we compare them to more exact results obtained through simulations in Appendix~\ref{appendix:forecasterrorsIMA} -- see in particular Fig.~\ref{fig:robustIMA}.  For $m > 30$ the approximation is excellent, but there are discrepancies for small values of $m$.

As before the right hand side is independent of all the parameters of the technology as well as the time horizon.  Eq.~(\ref{eq:distrtMA}) can be viewed as the distribution of errors around a point forecast, which makes it possible to collapse many technologies onto a single distribution.  This property is extremely useful for statistical testing, i.e. for determining the quality of the model.  But its greatest use, as we demonstrate in Section~\ref{comparison}, is that it makes it possible to formulate a distributional forecast for the future costs of a given technology.  

When $m$ is sufficiently large the Student $t$ distribution is well-approximated by a standard normal. Using the mean given by Eq.~(\ref{eq:predictionMRW}) and the variance determined by Eqs.~(\ref{eq:eIMAnorm}-\ref{eq:Astar}), the distributional forecast for the future logarithm of the cost $y_{t + \tau}$ conditioned on $(y_t, \ldots, y_{t-m+1})$ is\footnote{
Note that although we make the estimate of the variance $\theta$-dependent, we always use the estimate of the mean corresponding to $\theta = 0$.  We do this because this is simpler and more robust.}
\begin{equation}
y_{t+\tau} \sim \mathcal{N}(y_{t} + \hat{\mu} \tau, \hat{K}^2 A^*/(1+\theta^2)).
\label{eq:distforecast}
\end{equation}
We will return later to the estimation of $\theta$.

\subsection{Alternative hypotheses}

In addition to autocorrelation we investigated other ways to generalize the model, such as heavy tails and long-memory.  As discussed in Appendix~\ref{section:heavytailinnovations}, based on forecast errors we found little evidence for heavy tails.  Long-memory is in a sense an extreme version of the autocorrelation hypothesis\footnote{
A process has long-memory if the autocorrelation function of its increments is not integrable.  Under the long-memory hypothesis one expects the diffusion term of the normalized squared errors to scale as $\Xi (\tau) \sim \tau^{2H}$, where $H$ is the Hurst exponent.   In the absence of long-memory $H = 1/2$, but for long-memory $1/2 < H < 1$.   Long-memory can arise from many causes, including nonstationarity.  It is easy to construct plausible processes with the $\mu$ parameter varying where the mean squared errors grow faster than $\tau^2$.},
which produces errors that grow faster as a function of the forecasting horizon $\tau$ than a random walk.  Given that long-memory is a natural result of nonstationarity, which is commonly associated with technological change, our prior was that it was a highly plausible alternative.  However, as we will see, the geometric random walk with normal noise increments and autocorrelations seems to give good agreement for the time scaling of forecasting errors, so we did not investigate long-memory further.

\section{Data\label{data}}

\subsection{Data collection}

\begin{figure}[H]
\centering
\includegraphics[height=70mm]{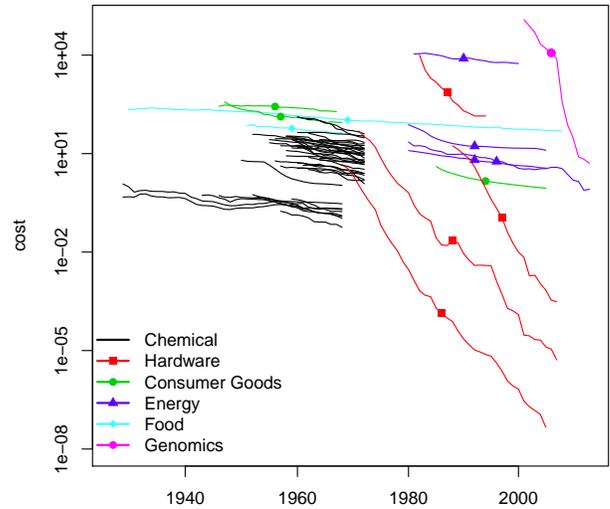}
\caption{{\it Cost vs. time for each technology in our dataset.}  This shows the 53 technologies out of the original set of 66 that have a significant rate of cost improvement (DNA sequencing is divided by 1000 to fit on the plot; the $y$-axis is in log scale).   More details can be found in Table \ref{table:descriptive} and Appendix \ref{appendix:data}.}
\label{fig:data}
\end{figure}

The bulk of our data on technology costs comes from the Santa Fe Institute's Performance Curve DataBase\footnote{pcdb.santafe.edu},  which was originally developed by Bela Nagy and collaborators; we augment it with a few other datasets.  These data were collected via literature search, with the principal criterion for selection being availability.  Fig.~\ref{fig:data} plots the time series for each data set.  The motley character of our dataset is clear: The time series for different technologies are of different lengths and they start and stop at different times. The sharp cutoff for the chemical data, for example, reflects the fact that it comes from a book published by the Boston Consulting Group in \citeyear{BSG}.  Table \ref{table:descriptive} gives a summary of the properties of the data and more description of the sources can be found in Appendix~\ref{appendix:data}.  This plot also makes it clear that technologies improve at very different rates.

A ubiquitous problem in forecasting technological progress is finding invariant units.   A favorable example is electricity.   The cost of generating electricity can be measured in dollars per kWh, making it possible to sensibly compare competing technologies and measure their progress through time.  Even in this favorable example, however, making electricity cleaner and safer has a cost, which has affected historical prices for technologies such as coal and nuclear power in recent years, and means that their costs are difficult to compare to clean and safe but intermittent sources of power such as solar energy.  To take an unfavorable example, our dataset contains appliances such as television sets, that have dramatically increased in quality through time\footnote{
\citet{gordon1990measurement} provides quality change adjustments for a number of durable goods. These methods (typically hedonic regressions) require additional data.}.  
Yet another problem is that some of them are potentially subject to scarcity constraints, which might potentially introduce additional trends and fluctuations.

One should therefore regard our results here as a lower bound on what is possible, in the sense that performing the analysis with better data in which all technologies had invariant units would very likely improve the quality of the forecasts.  We would love to be able to make appropriate normalizations but the work involved is prohibitive; if we dropped all questionable examples we would end with little remaining data.   Most of the data are costs, but in a few cases they are prices; again, this adds noise but if we were able to be consistent that should only improve our results.  We have done various tests removing data and the basic results are not sensitive to what is included and what is omitted (see Fig. \ref{fig:robustNdataset} in the appendix).

We have removed some technologies that are too similar to each other from the Performance Curve Database. For instance, when we have two datasets for the same technology, we keep only one of them. Our choice was based on data quality and length of the time series. This selection left us with 66 technologies belonging to different sectors that we label as chemistry, genomics, energy, hardware, consumer durables and food.

\subsection{Data selection and descriptive statistics}

In this paper we are interested in technologies that are improving, so we restrict our analysis to those technologies whose rate of improvement is statistically significant based on the available sample.  We used a simple one-sided $t$-test on the first-difference (log) series and removed all technologies for which the $p$-value indicates that we can't reject the null that $\mu_j =0$ at a 10\% confidence level\footnote{
This is under the assumption that $\theta = 0$.}.

\begin{table}
\vspace{-2cm}
\centering
\scalebox{0.62}{
\begin{tabular}{|c|c|c|c|c|c|c|c|}
  \hline
\rule{0pt}{14pt} 
Technology & Industry & T & $\tilde{\mu}$ & $p$ value & $\tilde{K}$ & $\tilde{\theta}$ \\ 
  \hline
Transistor & Hardware & 38 & -0.50 & 0.00 & 0.24 & 0.19 \\ 
  Geothermal.Electricity & Energy & 26 & -0.05 & 0.00 & 0.02 & 0.15 \\ 
  Milk..US. & Food & 79 & -0.02 & 0.00 & 0.02 & 0.04 \\ 
  DRAM & Hardware & 37 & -0.45 & 0.00 & 0.38 & 0.14 \\ 
  Hard.Disk.Drive & Hardware & 20 & -0.58 & 0.00 & 0.32 & -0.15 \\ 
  Automotive..US. & Cons. Goods & 21 & -0.08 & 0.00 & 0.05 & 1.00 \\ 
  Low.Density.Polyethylene & Chemical & 17 & -0.10 & 0.00 & 0.06 & 0.46 \\ 
  Polyvinylchloride & Chemical & 23 & -0.07 & 0.00 & 0.06 & 0.32 \\ 
  Ethanolamine & Chemical & 18 & -0.06 & 0.00 & 0.04 & 0.36 \\ 
  Concentrating.Solar & Energy & 26 & -0.07 & 0.00 & 0.07 & 0.91 \\ 
  AcrylicFiber & Chemical & 13 & -0.10 & 0.00 & 0.06 & 0.02 \\ 
  Styrene & Chemical & 15 & -0.07 & 0.00 & 0.05 & 0.74 \\ 
  Titanium.Sponge & Chemical & 19 & -0.10 & 0.00 & 0.10 & 0.61 \\ 
  VinylChloride & Chemical & 11 & -0.08 & 0.00 & 0.05 & -0.22 \\ 
  Photovoltaics & Energy & 34 & -0.10 & 0.00 & 0.15 & 0.05 \\ 
  PolyethyleneHD & Chemical & 15 & -0.09 & 0.00 & 0.08 & 0.12 \\ 
  VinylAcetate & Chemical & 13 & -0.08 & 0.00 & 0.06 & 0.33 \\ 
  Cyclohexane & Chemical & 17 & -0.05 & 0.00 & 0.05 & 0.38 \\ 
  BisphenolA & Chemical & 14 & -0.06 & 0.00 & 0.05 & -0.03 \\ 
  Monochrome.Television & Cons. Goods & 22 & -0.07 & 0.00 & 0.08 & 0.02 \\ 
  PolyethyleneLD & Chemical & 15 & -0.08 & 0.00 & 0.08 & 0.88 \\ 
  Laser.Diode & Hardware & 13 & -0.36 & 0.00 & 0.29 & 0.37 \\ 
  PolyesterFiber & Chemical & 13 & -0.12 & 0.00 & 0.10 & -0.16 \\ 
  Caprolactam & Chemical & 11 & -0.10 & 0.00 & 0.08 & 0.40 \\ 
  IsopropylAlcohol & Chemical & 9 & -0.04 & 0.00 & 0.02 & -0.24 \\ 
  Polystyrene & Chemical & 26 & -0.06 & 0.00 & 0.09 & -0.04 \\ 
  Polypropylene & Chemical & 10 & -0.10 & 0.00 & 0.07 & 0.26 \\ 
  Pentaerythritol & Chemical & 21 & -0.05 & 0.00 & 0.07 & 0.30 \\ 
  Ethylene & Chemical & 13 & -0.06 & 0.00 & 0.06 & -0.26 \\ 
  Wind.Turbine..Denmark. & Energy & 20 & -0.04 & 0.00 & 0.05 & 0.75 \\ 
  Paraxylene & Chemical & 12 & -0.10 & 0.00 & 0.09 & -1.00 \\ 
  DNA.Sequencing & Genomics & 13 & -0.84 & 0.00 & 0.83 & 0.26 \\ 
  NeopreneRubber & Chemical & 13 & -0.02 & 0.00 & 0.02 & 0.83 \\ 
  Formaldehyde & Chemical & 11 & -0.07 & 0.00 & 0.06 & 0.36 \\ 
  SodiumChlorate & Chemical & 15 & -0.03 & 0.00 & 0.04 & 0.85 \\ 
  Phenol & Chemical & 14 & -0.08 & 0.00 & 0.09 & -1.00 \\ 
  Acrylonitrile & Chemical & 14 & -0.08 & 0.01 & 0.11 & 1.00 \\ 
  Beer..Japan. & Food & 18 & -0.03 & 0.01 & 0.05 & -1.00 \\ 
  Primary.Magnesium & Chemical & 40 & -0.04 & 0.01 & 0.09 & 0.24 \\ 
  Ammonia & Chemical & 13 & -0.07 & 0.02 & 0.10 & 1.00 \\ 
  Aniline & Chemical & 12 & -0.07 & 0.02 & 0.10 & 0.75 \\ 
  Benzene & Chemical & 17 & -0.05 & 0.02 & 0.09 & -0.10 \\ 
  Sodium & Chemical & 16 & -0.01 & 0.02 & 0.02 & 0.42 \\ 
  Methanol & Chemical & 16 & -0.08 & 0.02 & 0.14 & 0.29 \\ 
  MaleicAnhydride & Chemical & 14 & -0.07 & 0.03 & 0.11 & 0.73 \\ 
  Urea & Chemical & 12 & -0.06 & 0.03 & 0.09 & 0.04 \\ 
  Electric.Range & Cons. Goods & 22 & -0.02 & 0.03 & 0.04 & -0.14 \\ 
  PhthalicAnhydride & Chemical & 18 & -0.08 & 0.03 & 0.15 & 0.31 \\ 
  CarbonBlack & Chemical & 9 & -0.01 & 0.03 & 0.02 & -1.00 \\ 
  Titanium.Dioxide & Chemical & 9 & -0.04 & 0.04 & 0.05 & -0.41 \\ 
  Primary.Aluminum & Chemical & 40 & -0.02 & 0.06 & 0.08 & 0.39 \\ 
  Sorbitol & Chemical & 8 & -0.03 & 0.06 & 0.05 & -1.00 \\ 
  Aluminum & Chemical & 17 & -0.02 & 0.09 & 0.04 & 0.73 \\ 
\hline 
  Free.Standing.Gas.Range & Cons. Goods & 22 & -0.01 & 0.10 & 0.04 & -0.30 \\
  CarbonDisulfide & Chemical & 10 & -0.03 & 0.12 & 0.06 & -0.04 \\ 
  Ethanol..Brazil. & Energy & 25 & -0.05 & 0.13 & 0.22 & -0.62 \\ 
  Refined.Cane.Sugar & Food & 34 & -0.01 & 0.23 & 0.06 & -1.00 \\ 
  CCGT.Power & Energy & 10 & -0.04 & 0.25 & 0.15 & -1.00 \\ 
  HydrofluoricAcid & Chemical & 11 & -0.01 & 0.25 & 0.04 & 0.13 \\ 
  SodiumHydrosulfite & Chemical & 9 & -0.01 & 0.29 & 0.07 & -1.00 \\ 
  Corn..US. & Food & 34 & -0.02 & 0.30 & 0.17 & -1.00 \\ 
  Onshore.Gas.Pipeline & Energy & 14 & -0.02 & 0.31 & 0.14 & 0.62 \\ 
  Motor.Gasoline & Energy & 23 & -0.00 & 0.47 & 0.05 & 0.43 \\ 
  Magnesium & Chemical & 19 & -0.00 & 0.47 & 0.04 & 0.58 \\ 
  Crude.Oil & Energy & 23 & 0.01 & 0.66 & 0.07 & 0.63 \\ 
  Nuclear.Electricity & Energy & 20 & 0.13 & 0.99 & 0.22 & -0.13 \\ 
   \hline
\end{tabular}
}
\caption{Descriptive statistics and parameter estimates (using the full sample) for all available technologies.  They are ordered by the $p$-value of a one-sided $t$-test for $\tilde{\mu}$, i.e. based on how strong the evidence is that they are improving.  The improvement of the last 13 technologies is not statistically significant and so they are dropped from further analysis -- see the discussion in the text. }   
\label{table:descriptive}
\end{table}

\begin{figure}[H]
\centering
\includegraphics[height=70mm]{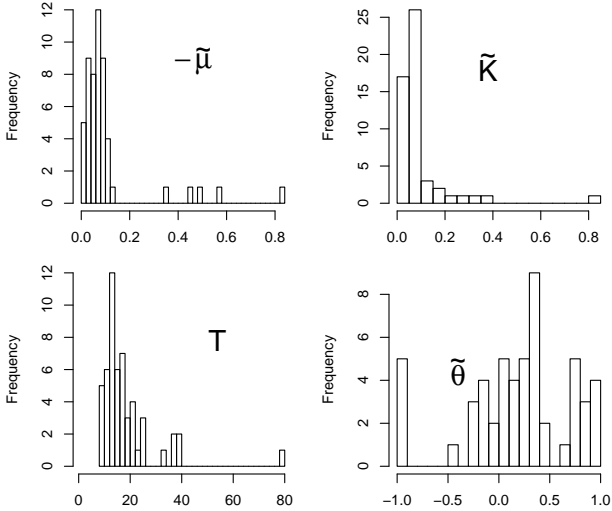}
\caption{{\it Histogram for the estimated parameters} for each technology $i$ based on the full sample (see also Table \ref{table:descriptive}). $\tilde{\mu}_j$ is the annual logarithmic rate of decrease in cost, $\tilde{K}_j$ is the standard deviation of the noise, $T_j$ is the number of available years of data and $\tilde{\theta}_j$ is the autocorrelation.}
\label{fig:histpara}
\end{figure}

Table \ref{table:descriptive} reports the $p$-values for the one sided $t$-tests and the bottom of the table shows the technologies that are excluded as a result. Table \ref{table:descriptive} also shows the estimated drift $\tilde{\mu}_j$ and the estimated standard deviation $\tilde{K_j}$ based on the full sample for each technology $j$.  (Throughout the paper we use a hat to denote estimates performed within an estimation window of size $m$ and a tilde to denote the estimates made using the full sample). Histograms of $\tilde{\mu}_j$, $\tilde{K}_j$, sample size $T_j$ and $\tilde{\theta}_j$ are given\footnote{The $\tilde{\theta}_j$ are estimated by maximum likelihood letting $\hat{\mu}_{MLE}$ be different from $\hat{\mu}$.} in Fig.~\ref{fig:histpara}.

\subsection{Relation between drift and volatility}

\begin{figure}[H]
\centering
\includegraphics[height=70mm]{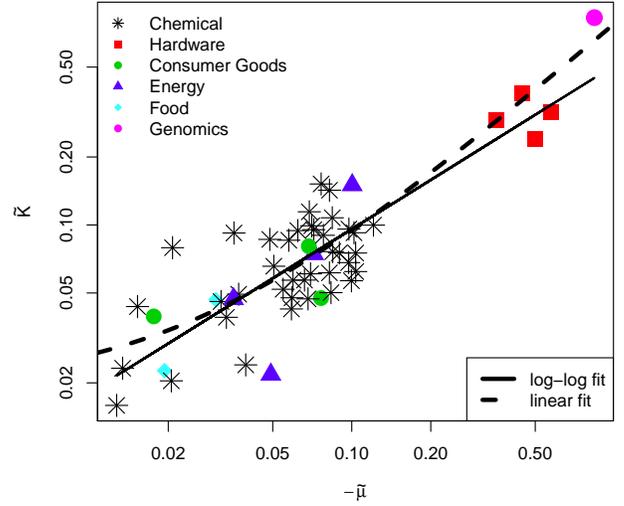}
\caption{{\it Scatter plot of the estimated standard deviation $\tilde{K_j}$ for technology $j$ against its estimated improvement rate $-\tilde{\mu}_j$}.  The dashed line shows a linear fit (which is curved when represented in log scale); the solid line is a log-log fit.  Technologies with a faster rate of improvement have higher uncertainty in their improvement.}
\label{fig:muvsk}
\end{figure}

Fig.~\ref{fig:muvsk} shows a scatter plot of the estimated standard deviation $\tilde{K}_j$ for technology $j$ vs. the estimated improvement rate $-\tilde{\mu}_j$.  A linear fit gives $\tilde{K} = 0.02 - 0.76 \tilde{\mu}$ with $R^2=0.87$ and standard errors of $0.008$ for the intercept and $0.04$ for the slope, as shown in the figure.  A log-log fit gives $\tilde{K} = e^{-0.68}(-\tilde{\mu})^{ 0.72}$ with 
$R^2=0.73$ and standard errors for the scaling constant of $0.18$ and for the exponent of $0.06$.  This indicates that on average the uncertainty $\tilde{K}_j$ gets bigger as the improvement rate $-\tilde{\mu_j}$ increases.  There is no reason that we are aware of to expect this {\it a priori}.  One possible interpretation is that for technological investment there is a trade-off between risk and returns. Another possibility is that faster improvement amplifies fluctuations. 

\section{Estimation procedures}
\label{section:estim}
\subsection{Statistical validation}
\label{section:hindcasting}

We use hindcasting for statistical validation, i.e. for each technology we pretend to be at a given date in the past and make forecasts for dates in the future relative to the chosen date\footnote{
This method is also sometimes called backtesting and is a form of cross-validation.}.
We have chosen this procedure for several reasons.  First, it directly tests the predictive power of the model rather than its goodness of fit to the data, and so is resistant to overfitting.  Second, it mimics the same procedure that one would follow in making real predictions, and third, it makes efficient use of the data available for testing.
 
We fit the model at each time step to the $m$ most recent changes in cost (i.e. the most recent $m+1$ years of data).  We use the same value of $m$ for all technologies and for all forecasts.  Because most of the time series in our dataset are quite short, and because we are more concerned here with testing the procedure we have developed rather than with making optimal forecasts, unless otherwise noted we choose $m = 5$.  This is admittedly very small, but it has the advantage that it allows us to make a large number of forecasts.  We will return later to discuss the question of which value of $m$ makes the best forecasts.

We perform hindcasting exhaustively in the sense that we make as many forecasts as possible given the choice of $m$.   For technology $j$, the cost data $y_{t} = \log p_{t}$ exists in years $t = 1, 2, \ldots, T_j$.  We then make forecasts for each feasible year and each feasible time horizon, i.e. we make forecasts $\hat{y}_{t_0 + \tau}(t_0)$ rooted in years $t_0 = (m+1, \ldots, T_j -1)$ with forecast horizon $\tau = (1, \ldots, T_j - t_0 )$. 

\begin{figure}[H]
 \centering
 \includegraphics[height=70mm]{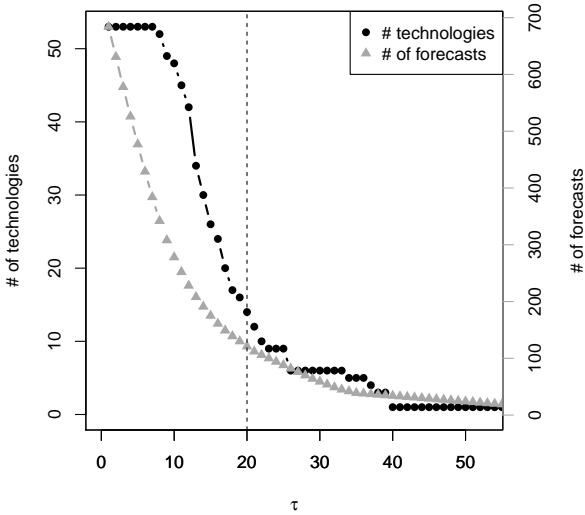}
 \caption{{\it Data available for testing as a function of the forecast time horizon.}  Here {\it \# of technologies} refers to the number of technology time series that are long enough to make at least one forecast at a given time horizon $\tau$, which is measured in years.   Similarly {\it \# of forecasts} refers to the total number of forecasts that can be made at time horizon $\tau$.   The horizontal line at $\tau=20$ years indicates our (somewhat arbitrary) choice of a maximum time horizon.  }
 \label{fig:numerror}
 \end{figure}
 
Since our dataset includes technology time series of different length (see Table~\ref{table:descriptive} and Fig.~\ref{fig:data}) the number of possible forecasts that can be made with a given historical window $m$ is highest for $\tau = 1$ and decreases for longer horizons\footnote{
The number of possible forecasts that can be made using a technology time series of length $T_j$ is $[T_j-(m+1)][T_j-m]/2$ which is $O(T_j^2).$ Hence the total number of forecast errors contributed by a given technology time series is disproportionately dependent on its length. However, we have checked that aggregating the forecast errors so that each technology has an equal weight does not qualitatively change the results.
}.
Fig.~\ref{fig:numerror} shows the total number of possible forecasts that can be made with our dataset at a given horizon $\tau$  and the number of technology time series that are long enough to make at least one forecast at horizon $\tau$.  This shows that the amount of available data decreases dramatically for large forecast horizons.  We somewhat arbitrarily impose an upper bound of $\tau_{max}=20$, but find this makes very little difference in the results (see Appendix \ref{section:appendixtaumax}).  There are a total of 8212 possible forecasts that can be made with an historical window of $m = 5$, and 6391 forecasts that can be made with $\tau \leq 20$.

To test for statistical significance we use a surrogate data procedure (explained below).  There are three reasons for doing this:  The first is that, although we derived approximate formulas for the forecast errors in Eq.~(\ref{MSFEmainIMA}) and (\ref{eq:distrtMA}), when $\theta\neq0$ the approximation is not very good for $m=5$. The second is that the rolling window approach we use for hindcasting implies overlaps in both the historical sample used to estimate parameters at each time $t_0$ and overlapping intervals in the future for horizons with $\tau > 1$. This implies substantial correlation in the empirical forecast errors, which complicates statistical testing.  The third reason is that, even if the formulas were exact, we expect finite sample fluctuations.  That is, with a limited number of technologies and short time series, we do not expect to find the predicted result exactly; the question is then whether the deviation that we observe is consistent with what is expected.
  
The surrogate data procedure estimates a null distribution for the normalized mean squared forecast error under the hypothesized model.  This is done by simulating both the model and the forecasting procedure to create a replica of the dataset and the forecasts. This is repeated for many different realizations of the noise process in order to generate the null distribution.  More specifically, for each technology we generate $T_j$ pseudo cost data points using Eq. (\ref{eq:IMA11}) with $\mu = \tilde{\mu}_j$, $K = \tilde{K}_j$ and a given value of $\theta$, thereby mimicking the structure of the data set.  We then estimate the parameters and perform hindcasting just as we did for the real data, generating the same number of forecasts and computing the mean squared forecast error.  This process is then repeated many times with different random number seeds to estimate the distribution.  This same method can be used to estimate expected deviations for any quantity, e.g. we also use this to estimate the expected deviation of the finite sample distribution from the predicted distribution of forecast errors.

\subsection{Parameter estimation \label{parameterEstimation}} 

We estimate the mean and the variance for each technology dynamically, using a rolling window approach to fit the parameters based on the $m + 1$ most recent data points.  In each year $t_0$ for which forecasts are made the drift $\hat{\mu}_{t_0}$ is estimated as the sample mean of the first differences,
\begin{equation} 
\hat{\mu}_{t_0} =\frac{1}{m}\sum_{i=t_0-m}^{t_0-1}(y_{i+1}-y_i)=\frac{y_{t_0}-y_{t_0 - m}}{m},
\label{eq:muhat0}
\end{equation} 
where the last equality follows from the fact that the sum is telescopic, and implies that only two points are needed to estimate the drift. The volatility is estimated using the unbiased estimator\footnote{This is different from the maximum likelihood estimator, which does not make use of Bessel's correction (i.e. dividing by ($m-1$) instead of $m$). Our choice is driven by the fact that in practice we use a very small $m$, making the bias of the maximum likelihood estimator rather large.}
\begin{equation}
\hat{K}^2_{t_0}=\frac{1}{m-1}\sum_{i=t_0 - m  }^{t_0-1}[(y_{i+1}-y_i)-\hat{\mu}_{t_0}]^2.
\end{equation}
This procedure gives us a variable number of forecasts for each technology $j$ and time horizon $\tau$ rooted at all feasible times $t_0$. We record the forecasting errors $\mathcal{E}_{t_0,\tau} = y_{t + \tau}(t_0)  - \hat{y}_{t + \tau}(t_0)$ and the associated values of $\hat{K}_{t_0}$ for all $t_0$ and all $\tau$ where we can make forecasts.

The autocorrelation parameter $\theta$ for the generalized model has to be treated differently.  Our time series are simply too short to make reasonable rolling window, technology-specific estimates for $\theta$.  With such small values of $m$ the estimated autocorrelations are highly unreliable.  

Our solution is to use a {\it global} value of $\theta$, i.e. we use the same value for all technologies and all points in time.  It may well be that $\theta$ is technology specific, but given the short amount of data it is necessary to make a choice that performs well under forecasting.  This is a classic bias-variance trade-off, where the variance introduced by statistical estimation of a parameter is so large that the forecasts produced by a biased model with this parameter fixed are superior. With very long time series this could potentially be avoided.  This procedure seems to work well. It leaves us with a parameter that has to be estimated in-sample, but since this is only one parameter estimated from a sample of more than $6,000$ forecasts the resulting estimate should be reasonably reliable. 

Evidence concerning autocorrelations is given in Fig.~\ref{fig:histpara}, where we present a histogram for the values of $\tilde{\theta}_j$ for each technology $j$ based on the full sample.  The results are highly variable.  Excluding eight likely outliers where $\tilde{\theta}_j = \pm 1$, the mean across the sample is $0.27$, and 35 out of the remaining 45 improving technologies have positive values of $\tilde{\theta}_j$.  This seems to suggest that $\theta$ tends to be positive.

We use two different methods for estimating a global value of $\theta$.  The first method takes advantage of the fact that the magnitude of the forecast errors is an increasing function of $\theta$ (we assume $\theta>0$) and chooses $\theta_{m}$ ($m$ as in ``matched'') to match the empirically observed forecast errors, leading to $\theta_m=0.63$ as described in the next section.  The second method takes a weighted average $\theta_w$ ($w$ as in ``weighted'') calculated as follows. We exclude all technologies for which the estimate of $\theta$ reveals specification or estimation issues ($\theta \approx 1$ or $\theta \approx -1$). Then at each horizon we compute a weighted average, with the weights proportional to the number of forecasts made with that technology.  Finally we take the average of the first 20 horizon-specific estimated values of $\theta$, leading to $\theta_w=0.25$. See Appendix \ref{section:thetaselection}.

\section{Comparison of models to data\label{empiricalResults}}
\label{section:results}

In comparing the model to data we address the following five questions:
\begin{enumerate}
\item
Is the scaling law for the increase in forecasting errors as a function of time derived in Eqs.~(\ref{MSFEmain}) and (\ref{MSFEmainIMA}) consistent with the data?
\item
Does there exist a value of $\theta$ such that the null hypothesis of the model is not rejected?  If so, what is this value, and how strong is the evidence that it is positive?
\item
When the normalized errors for different technologies at different time horizons are collapsed onto a single distribution, does this agree with the Student distribution as predicted by Eq.~(\ref{eq:distrtMA})?
\item
Do the errors scale with the trailing sample size $m$ as predicted under the assumption that the random process is stationary (i.e. that parameters are not changing in time)?
\item
Is the model well-specified?
\end{enumerate}
We will see that we get clear affirmative answers to the first four questions but we are unable to answer question (5).

\subsection{Normalized forecast errors as a function of $\tau$}

\begin{figure}[h]
\centering
\includegraphics[height=70mm]{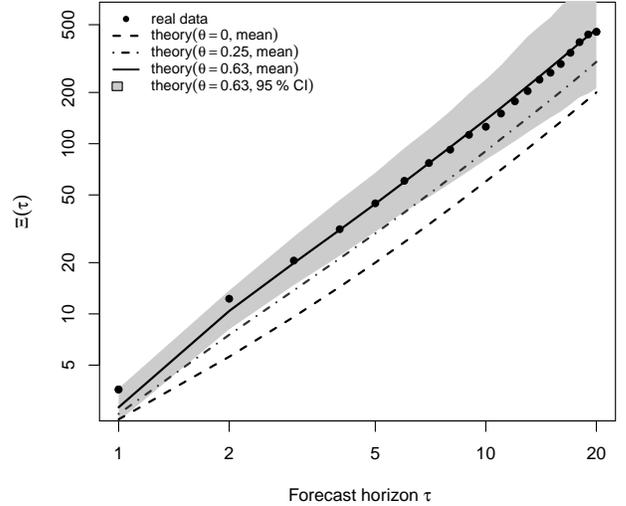}
\caption{{\it Growth of the mean squared normalized forecast error $\Xi(\tau)$ for the empirical forecasts compared to predictions using different values of $\theta$}.   The empirical value of the normalised error $\Xi(\tau)$ is shown by black dots.  The grey area corresponds to the $95\%$ confidence intervals for the case $\theta=\theta_{m}$.  The dashed line represents the predicted squared normalized error with $\theta = 0$, the dot-dash line is for $\theta_w = 0.25$ and the solid line is for $\theta_{m} = 0.63$.}
\label{fig:errorgrowth}
\end{figure}

To answer the first question we compute the sample estimate of the mean squared normalized forecast error $\Xi (\tau)$, averaging over all available forecasts for all technologies at each time horizon with $\tau \le 20$ (see Eq.~(\ref{MSFEmainIMA})). Fig.~\ref{fig:errorgrowth} compares the empirical results to the model with three different values of the autocorrelation parameter $\theta$. 
Because the approximate error estimates derived in Eq.~(\ref{MSFEmainIMA}) break down for small values of $m$, for each value of $\theta$ we estimate the expected mean squared errors under the null hypothesis of the model via the surrogate data procedure described in Section~\ref{section:hindcasting}\footnote{When $\theta=0$ the simulated and analytical results are visually indistiguishable. Fig.~\ref{fig:errorgrowth} uses the analytical formula, Eq.~(\ref{MSFEmain}).}.

The model does a good job of predicting the scaling of the forecast errors as a function of the time horizon $\tau$. The errors are predicted to grow approximately proportional to $(\tau + \tau^2/m)$; at long horizons the error growth at each value of $\theta$ closely parallels that for the empirical forecasts. This suggests that this scaling is correct, and that there is no strong support for modifications such as long-memory that would predict alternative rates of error growth.

Using $\theta_{m} = 0.63$ gives a good match to the empirical data across the entire range of time horizons.  Note that even though we chose $\theta_m$ in order to get the best possible match, given that we are rescaling data for different technologies by the empirically measured sample standard deviations over very short samples of length $m = 5$, and that we are predicting across 20 different time horizons simultaneously, the ability to find a value of the parameter $\theta$ that matches this well was far from guaranteed. (It is completely possible, for example, that there would simply not exist a value of $\theta < 1$ yielding errors that were sufficiently large).  

To test the statistical significance of the results for different values of $\theta$ and $\tau$ we use the surrogate data procedure described at the end of Section~\ref{section:hindcasting}. For $\theta_m = 0.63$ we indicate error bars by showing in grey the region containing the 95\% of the simulated realizations with errors closest to the mean. For $\tau = 1$ and $\tau = 2$ the predicted errors are visibly below the empirical observations, but the difference is within the error bars (though on the edge of the error bars for $\tau =1$); the agreement is very good at all other values of $\tau$.  The autocorrelation parameter $\theta_w = 0.25$ is weakly rejected for $\tau$ between 1 and 6 and weakly accepted elsewhere, indicating that it is very roughly the lowest value of $\theta$ that is consistent with the data at the two standard deviation level. In contrast the case $\theta = 0$, which gives normalized error predictions that are lower by about a factor of two, is clearly well outside of the error bars (note the logarithmic scale).  This strongly indicates that a positive value of $\theta$ is required to match the observed errors, satisfying $\theta > \theta_w  = 0.25$.

\subsection{Distribution of forecast errors}

\begin{figure}[H]
\includegraphics[height=70mm]{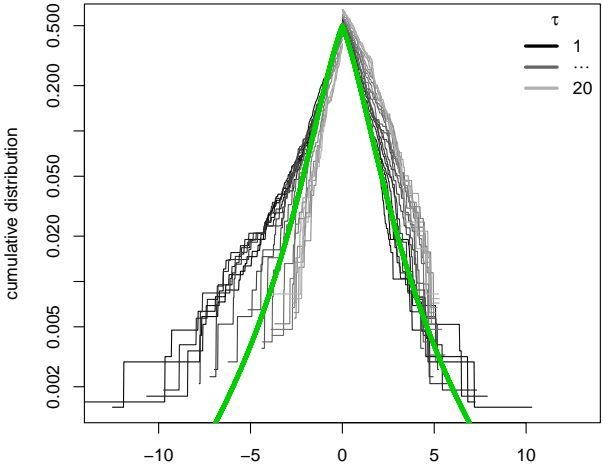}
\caption{{\it Cumulative distribution of empirical rescaled normalized forecast errors at different forecast horizons $\tau$}. The forecast errors for each technology $j$ are collapsed using Eq.~(\ref{eq:distrtMA}) with $\theta = \theta_{m} = 0.63$. This is done for each forecast horizon $\tau = 1, 2, \ldots, 20$ as indicated in the legend.  The green thick curve is the theoretical prediction.  The positive and negative errors are plotted separately.  For the positive errors we compute the number of errors greater than a given value $X$ and divide by the total number of errors to estimate the cumulative probability and plot in semi-log scale.  For the negative errors we do the same except that we take the absolute value of the error and plot against $-X$.}
\label{fig:tailplot}
\end{figure}

We now address question (3) by testing whether we correctly predict the distribution of forecast errors.  Fig.~\ref{fig:tailplot} shows the distribution of rescaled forecast errors using $\theta_{m} = 0.63$ with Eq.~(\ref{eq:distrtMA}) to rescale the errors.  Different values of $\tau$ are plotted separately, and each is compared to the predicted Student distribution. Overall, the fit is good but at longer horizons forecast errors tend to be positive, that is, realized technological progress is slightly slower than predicted. We have tested to see if this forecast bias is significant, and for $\tau\leq11$ we cannot reject the null that there is no bias even at the $10\%$ level. At higher horizons there is evidence of forecast bias, but we have to remember that at these horizons we have much less data (and fewer technologies) available for testing.

\begin{figure}[H]
\includegraphics[height=70mm]{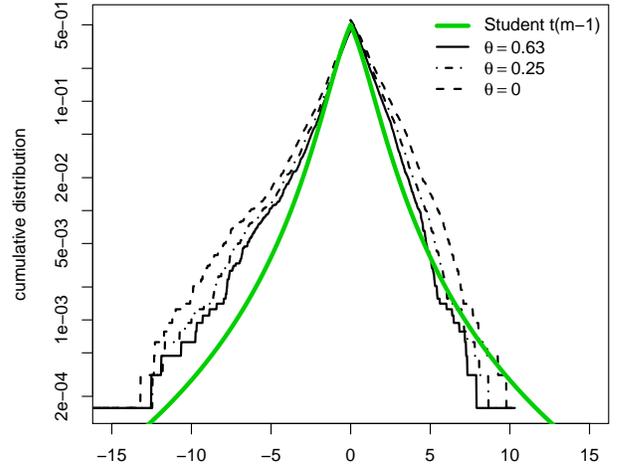}
\caption{{\it Cumulative distribution of empirical rescaled normalized forecast errors with all $\tau$ pooled together} for three different values of the autocorrelation parameter, $\theta = 0$ (dashed line), $\theta = 0.25$ (dot-dash line) and $\theta = 0.63$ (solid line).  See the caption of Fig.~\ref{fig:tailplot} for a description of how the cumulative distributions are computed and plotted.}
\label{fig:tailplotcompare}
\end{figure}

Fig.~\ref{fig:tailplotcompare} shows the empirical distribution with all values of $\tau$ pooled together, using rescalings corresponding to $\theta = 0$, $\theta_w$, and $\theta_m$.  The predicted distribution is fairly close to the theoretical prediction, and as expected the fit with $\theta_m  = 0.63$ is better than with $\theta_w = 0.25$ or $\theta = 0$.

To test whether the observed deviations of the empirical error distribution from the predicted distribution are significant  we once again use the surrogate data approach described at the end of Section~\ref{section:hindcasting}. As before we generate many replicas of the dataset and forecasts. For each replica of the dataset and forecasts we compute a set of renormalized errors $\epsilon^*$ and construct their distribution.  We then measure the average distance between the surrogate distribution and the Student distribution as described in Appendix~\ref{sec:statTest}.   Repeating this process 10,000 times results in the sampling distribution of the deviations from the Student distribution under the null hypothesis that the model is correct.  We then compare this to the corresponding value of the average distance between the real data and the Student distribution, which gives us a $p$-value under the null hypothesis. We find that the model with $\theta_{m} = 0.63$ is accepted.  In contrast $\theta_w = 0.25$ is rejected with $p$-values ranging from $1\%$ to $0.1\%$, depending on the way in which the average distance is computed. The case with $\theta = 0$ is very strongly rejected.

These results make it clear that the positive autocorrelations are both statistically significant and important. The statistical testing shows that $\theta = 0.63$ provides a good estimate for the observed forecasting errors across a large range of time horizons, with normalized forecasting errors that are well-described by the Student distribution.

\subsection{Dependence on sample size $m$ \label{sec:m}}

So far we have used only a small fraction of the data to make each forecast.  The choice for the trailing sample of $m=5$  was for testing purposes, allowing us to generate a large number of forecasts and test our method for estimating their accuracy. 

We now address the question of the optimal value of $m$.   If the process is stationary in the sense that the parameters $(\mu, K, \theta)$ are constant, one should always use the largest possible value of $m$.  If the process is nonstationary, however, it can be advantageous to use a smaller value of $m$, or alternatively a weighted average that decays as it goes into the past.  How stationary is the process generating technology costs, and what is the best choice of $m$?  

\begin{figure}[H]
\includegraphics[height=70mm]{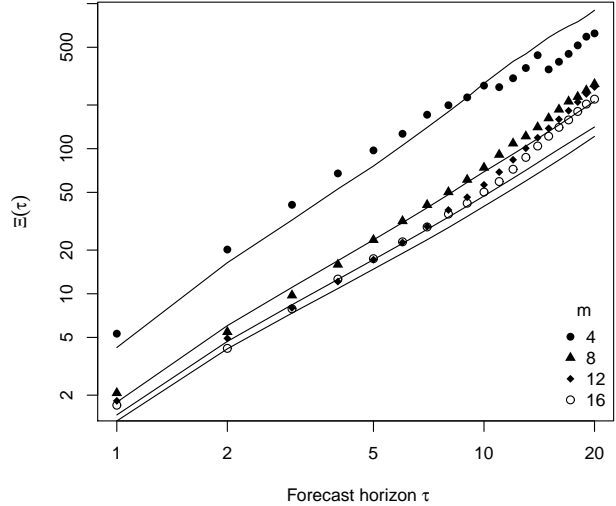}
\caption{\emph{Mean squared normalized forecast error $\Xi$ as a function of the forecast horizon $\tau$ for different sizes of the trailing sample size $m$.}  This is done for $m = (4, 8, 12, 16)$, as shown in the legend.   The corresponding theoretical predictions are made using $\theta_m = 0.63$, and are shown as solid curves ordered in the obvious way from top ($m = 4$) to bottom ($m = 16$).  }
\label{fig:errorgrowthrobustW}
\end{figure}

We experimented with increasing $m$, as shown in Fig.~\ref{fig:errorgrowthrobustW}, and compared this to the model with $\theta_m = 0.63$.  We find that the errors drop as $m$ increases roughly as one would expect if the process were stationary\footnote{
Note that to check forecast errors for high $m$ we have used only technologies for which at least $m+2$ years were available.  For large values of $m$ the statistical variation increases due to lack of data.}
and that the model does a reasonably good job of forecasting the errors (see also Appendix~\ref{section:robustm}).
This indicates that the best choice is the largest possible value of $m$, which in this case is $m = 16$.   However we should emphasize that it is entirely possible that testing on a sample with longer time series might yield an optimal value\footnote{
We present the results up to $m=16$ because less than a third of the technologies can be used with larger sample sizes. We have performed the same analysis up to $m=35$, where only 5 technologies are left, and the results remain qualitatively the same.}
of $m>16$.

\subsection{Is the model well-specified?}

Because most of our time series are so short it is difficult to say whether or not the model is well-specified.  As already noted, for such short series it is impossible to usefully estimate technology-specific values of the parameter $\theta$, which has forced us to use a global value for all technologies.  Averaging over the raw samples suggests a relatively low value $\theta_w = 0.25$, but a much higher value $\theta_m = 0.63$ is needed to match the empirically observed errors.
However we should emphasize that with such short series $\theta$ is poorly estimated, and it is not clear that averaging across different technologies is sufficient to fix this problem.

In our view it would be surprising if there are not technology-specific variations in $\theta$; after all $\mu_j$ and $K_j$ vary significantly across technologies.  So from this point of view it seems likely that the model with a global $\theta$ is mis-specified.  It is not clear whether this would be true if we were able to measure technology-specific values of $\theta_j$.  It is remarkable that such a simple model can represent a complicated process such as technological improvement as well as it does, and in any case, as we have shown, using $\theta = \theta_m$ does a good job of  matching the empirically observed forecasting errors.  Nonetheless, testing with more data is clearly desirable.

\section{Application to solar PV modules \label{comparison}}

In this section we provide a distributional forecast for the price of solar photovoltaic modules.  We then show how this can be used to make a comparison to a hypothetical competing technology in order to estimate the probability that one technology will be less expensive than another at a given time horizon.

\subsection{A distributional forecast for solar energy}

We have shown that the autocorrelated geometric random walk can be used to forecast technological cost improvement and that the formula we have derived for the distribution of forecast errors works well when applied to many different technologies.  We now demonstrate how this can be used to make a distributional forecast for the cost improvement of a given technology.  The fact that the method has been extensively tested on many technologies in the previous section gives us some confidence that this forecast is reliable.

\begin{figure}[H]
\includegraphics[height=70mm]{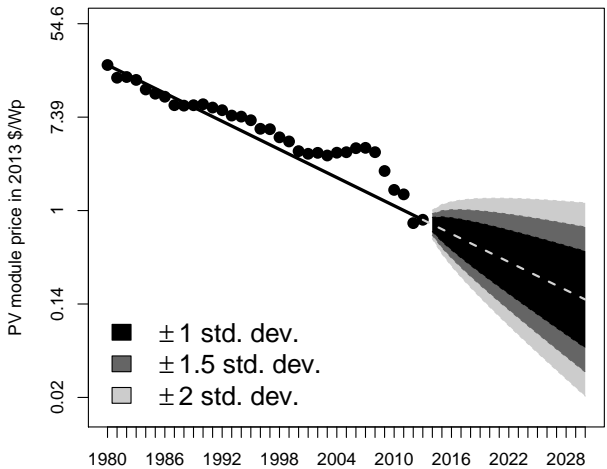}
\caption{\emph{Forecast for the cost of photovoltaic modules} in 2013 \$/\mbox{Wp}. 
The point forecasts and the error bars are produced using Eq.~(\ref{eq:distforecastsolar}) and the parameters discussed in the text.  Shading indicates the quantiles of the distribution corresponding to 1, 1.5 and 2 standard deviations.}
\label{fig:solarforecast}
\end{figure}

We make the forecast using Eq.~(\ref{eq:distforecast}).  We use all available years of past data ($m=33$) to fit the parameters $\hat{\mu}_S = \tilde{\mu}_S = -0.10$ and $\hat{K}_S = \tilde{K}_S = 0.15$, and we used $\theta = \theta_{m} = 0.63$.  The forecast is given by Eq.~(\ref{eq:distforecast}) with appropriate substitutions of parameters, i.e.
\begin{equation}
y_S(t+\tau) \sim \mathcal{N}(y_S(t)+\tilde{\mu}_S \tau,\tilde{K}_S^2 A^*/(1+{\theta_{m}}^2)),
\label{eq:distforecastsolar}
\end{equation}
where $A^*(\theta_m)$ is defined in Eq.~(\ref{eq:Astar}).  Fig.~\ref{fig:solarforecast} shows the predicted distribution of likely prices for solar photovoltaic modules for time horizons up to 2030.  The intervals corresponding to plus or minus two standard deviations in Fig. \ref{fig:solarforecast} are 95\% prediction intervals.

The prediction says that it is likely that solar PV modules will continue to drop in cost at the roughly $10\%$ rate that they have in the past.  Nonetheless there is a small probability (about $5\%$) that the price in 2030 will be higher than it was in 2013 \footnote{
This forecast is consistent with the one made several years ago by \cite{nagy2013statistical} using data only until 2009. It is difficult to compare this forecast with expert's elicitation studies, which are often more precise in terms of which PV technology and which market is predicted and are often concerned with levelized costs. Individual experts' distributional predictions for LCOE (see Fig. 6 in \citet{bosetti2012future}) seem tight as compared to ours (for modules only). However, the predictions for the probability that PV will cost less than \$0.30/Wp in 2030 reported in Fig.3 of \citet{curtright2008expert} are overall comparable with ours.}.
While it might seem remarkable to forecast 15 years ahead with only 33 years of past data, note that throughout most of the paper we were forecasting up to 20 years ahead with only six years of data.  As one uses more past data, the width of the distributional forecast decreases.  In addition there are considerable variations in the standard deviations $\tilde{K}_j$ of the technologies in Table~\ref{table:descriptive}; these variations are reflected in the width of the distribution at any given forecasting horizon.  The large deviation from the trend line that solar module costs made in the early part of the millennium cause the estimated future variance to be fairly large.  

Except for the estimation of $\theta$ no data from other technologies was used in this forecast.  Nonetheless,  data from other technologies were key in giving us confidence that the distributional forecast is reliable.

\subsection{Estimating the probability that one technology will be less expensive than another}
Suppose we want to compute the probability that a given technology will be less expensive than another competing technology at a given point in the future. We illustrate how this can be done by comparing the log cost of photovoltaic modules $y_S$ with the log cost of a hypothetical alternative technology $y_C$.   Both the cost of photovoltaic modules and technology C are assumed to follow Eq. (\ref{eq:distforecastsolar}), but for the sake of argument we assume that, like coal,  technology $C$ has historically on average had a constant cost, i.e. $\tilde{\mu}_C=0$. We also assume that the estimation period is the same, and that $\theta_C=\theta_S=\theta_{m}$.  We want to compute the probability that $\tau$ steps ahead $y_S<y_C$.  The probability that $y_S<y_C$ is the probability that the random variable $Z=y_C-y_S$ is positive. Since $y_S$ and $y_C$ are normal, assuming they are independent their difference is normal, i.e.
\[
Z \sim \mathcal{N}\left( \mu_Z, \sigma_Z^2 \right),
\]
where $\mu_Z=(y_C(t)-y_S(t))+ \tau (\tilde{\mu}_C -\tilde{\mu}_S)$ and $\sigma^2_Z=(A^*/(1+\theta_m^{2}))(\tilde{K}_S^2 + \tilde{K}_C^2)$. The probability that $y_S<y_C$ is the integral for the positive part, which is expressed in terms of the error function
\begin{equation}
\begin{split}
Pr(y_S<y_C)&=\int_0^\infty f_Z(z)dz\\
&=\frac{1}{2} \left[1 + \text{Erf}\left(\frac{\mu_Z}{\sqrt{2}\hspace{1mm} \sigma_Z} \right) \right].
\end{split}
\label{eq:erf}
\end{equation}

\begin{figure}[H]
\includegraphics[height=70mm]{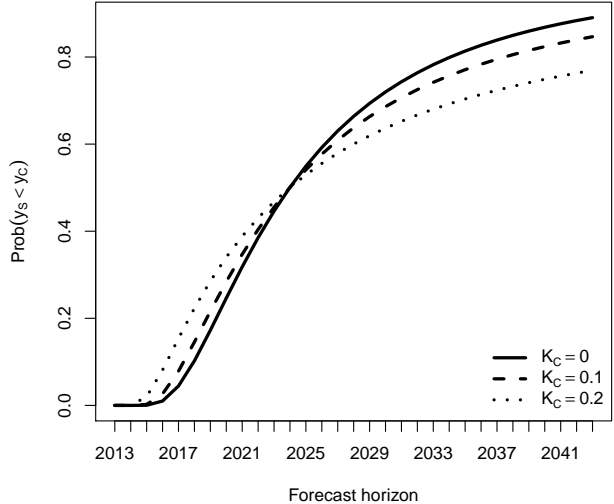}
\caption{\emph{Probability that solar photovoltaic modules become less expensive than a hypothetical competing technology $C$ whose initial cost is one third that of solar but is on average not improving,} i.e. $\tilde{\mu}_C = 0$. The curves show Eq.~(\ref{eq:erf}) using
 $\tilde{\mu}_S=-0.10$, $\tilde{K}_S=0.15$, $m=33$ for solar PV and three different values of the noise parameter $\tilde{K_C}$ for  technology $C$. The crossing point is at $\tau \approx 11$ (2024) in all three cases.}
\label{fig:psn}
\end{figure}

In Fig. \ref{fig:psn} we plot this function using the parameters estimated for photovoltaics, assuming that the cost of the competing technology is a third that of solar at the starting date in 2013, and that it is on average not dropping in cost, i.e. $\mu_C = 0$.  We consider three different levels of the noise parameter $\tilde{K}_C$ for technology $C$. Note that changing the noise parameter does not change the expected time when the curves cross. 

The main point of this discussion is that with our method we can reliably forecast the probability that a given technology will surpass a competitor.

\subsection{Discussion of PV relative to coal-fired electricity and nuclear power}

In the above discussion we have carefully avoided discussing a particular competing technology.   A forecast for the full cost of solar PV electricity requires predicting the balance of system costs, for which we lack consistent historical data, and unlike module costs, the full cost depends on factors such as insolation, interest rates and local installation costs.   As solar PV grows to be a significant portion of the energy supply the cost of storage will become very important.  Nonetheless, it is useful to discuss it in relation to the two competitors mentioned in the introduction.  

An analysis of coal-fired electricity, breaking down costs into their components and examining each of the trends separately, has been made by \cite{McNerney2011historical}.  They show that while coal plant costs (which are currently roughly $40\%$ of total cost) dropped historically, this trend reversed circa 1980.  Even if the recent trend reverses and plant construction cost drops dramatically in the future, the cost of coal is likely to eventually dominate the total cost of coal-fired electricity.   As mentioned before, this is because the historical cost of coal is consistent with a random walk without drift, and currently fuel is about $40\%$ of total costs.   If coal remains constant in cost (except for random fluctuations up or down) then this places a hard bound on how much the total cost of coal-fired electricity can decrease.  Since typical plants have efficiencies the order of $1/3$ there is not much room for making the burning of coal more efficient -- even a spectacular efficiency improvement to $2/3$ of the theoretical limit is only an improvement of a factor of two, corresponding to the average progress PV modules make in about 7.5 years.   Similar arguments apply to oil and natural gas\footnote{
Though much has been made of the recent drop in the price of natural gas due to fracking, which has had a large effect, one should bear in mind that the drop is tiny in comparison to the factor of about 2,330 by which solar PV modules have dropped in price.  The small change induced by fracking is only important because it is competing in a narrow price range with other fossil fuel technologies.  In work with other collaborators we have examined not just oil, coal and gas, but more than a hundred minerals; all of them show remarkably flat historical prices, i.e. they all change by less than an order of magnitude over the course of a century.}.

Because historical nuclear power costs have tended to increase, not just in the US but worldwide, even a forecast that they will remain constant seems optimistic.  Levelized costs for solar PV powerplants in 2013 were as low as 0.078-0.142 Euro/\mbox{kWh} (0.09-0.16\$) in Germany \citep{FISES2013}\footnote{
Levelized costs decrease more slowly than module costs, but do decrease \citep{nemet2006beyond}. For instance, installation costs per watt have fallen in Germany and are now about half what they are in the U.S. \citep{NREL2013}.}, 
and in 2014 solar PV reached a new record low with an accepted bid of \$0.06/\mbox{kWh} for a plant in Dubai\footnote{
See http://www.renewableenergyworld.com/rea/
news/article/2015/01/dubai-utility-dewa-
procures-the-worlds-cheapest-solar-energy-ever}. 
When these are compared to the projected cost of \$0.14/\mbox{kWh} in 2023 for the Hinkley Point nuclear reactor, it appears that the two technologies already have roughly equal costs, though of course a direct comparison is difficult due to factors such as intermittency, waste disposal, insurance costs, etc.

As a final note, skeptics have claimed that solar PV cannot be ramped up quickly enough to play a significant role in combatting global warming.  A simple trend extrapolation of the growth of solar energy (PV and solar thermal) suggests that it could represent 20\% of the energy consumption by 2027. In contrast the "hi-Ren" (high renewable) scenario of the International Energy Agency, which is presumably based on expert analysis, assumes that PV will generate $16\%$ of total electricity in 2050.  Thus even in their optimistic forecast they assume PV will take 25 years longer than the historical trend suggests (to hit a lower target).  We hope in the future to formulate similar methods for forecasting production so that we can better assess the reliability of such forecasts. See Appendix~\ref{solarAppendix} and Fig.~\ref{energyUsageComparison} in particular.

\section{Conclusion \label{conclusion}}

Many technologies follow a similar pattern of progress but with very different rates.  In this paper we have proposed a simple method based on the autocorrelated geometric random walk to provide robust predictions for technological progress that are stated as distributions of outcomes rather than point forecasts.  We assume that all technologies follow a similar process except for their rates of improvement and volatility.  Under this assumption we can pool forecast errors of different technologies to obtain an empirical estimation of the distribution of forecast errors.

One of the essential points of this paper is that the use of many technologies allows us to make a better forecast for a given technology, such as solar PV modules.  Although using many technologies does not affect our point forecast, it is the essential element that allowed us to test our distributional forecasts in order to ensure that they are reliable.  The point is that by treating all technologies as essentially the same except for their parameters, and collapsing all the data onto a single distribution, we can pool data from many technologies to gain confidence in and calibrate our method for a given technology.  It is of course a bold assumption to say that all technologies follow a random process with the same form, but the empirical results indicate that this a good hypothesis.  

We do not want to suggest in this paper that we think that Moore's law provides an optimal forecasting method.  Quite the contrary, we believe that by gathering more historical data, and by adding other auxiliary variables such as production, R\&D, patent activity, there should be considerable room for improving forecasting power.  In the future we anticipate that theories will eventually provide causal explanations for why technologies improve at such different rates and this will result in better forecasts.  Nonetheless, in the meantime the method we have introduced here provides a benchmark against which other approaches can be measured.  It provides a proof of principle that technologies can be successfully forecast and that the errors in the forecasts can be reliably predicted.  

From a policy perspective we believe that our method can be used to provide an objective point of comparison to expert forecasts, which are often biased by vested interests and other factors.  The fact that we can associate uncertainties with our predictions makes them far more useful than simple point forecasts.  The example of solar PV modules illustrates that differences in the improvement rate of competing technologies can be dramatic, and that an underdog can begin far behind the pack and quickly emerge as a front-runner. 
 Given the urgency of limiting greenhouse gas emissions, it is fortuitous that a green technology also happens to have such a rapid improvement rate, and is likely to eventually surpass its competition within $10 - 20$ years.  In a context where limited resources for technology investment constrain policy makers to focus on a few technologies that have a real chance to eventually achieve and even exceed grid parity, the ability to have improved forecasts and know how accurate they are should prove particularly useful.

\appendix
\section*{Appendix}

\section{Data}
\label{appendix:data}

The data are mostly taken from the Santa-Fe Performance Curve
DataBase, accessible at pcdb.santafe.edu. The database has been constructed
from personal communications and from \citet{colpier2002economics,goldemberg2004ethanol,lieberman1984learning,lipman1999experience,zhao1999diffusion,mcdonald2001learning,neij2003experience,moore2006behind,nemet2006beyond,schilling2009technology}. The data on photovoltaic prices has been collected from public releases of Strategies Unlimited, Navigant and SPV Market Research. The data on nuclear energy is from \citet{koomey2007reactor} and \citet{Cooper}. The DNA sequencing data is from \citet{dnadata} (cost per human-size genome), and for each year we took the last available month (September for 2001-2002 and October afterwards) and corrected for inflation using the US GDP deflator.

\section{Distribution of forecast errors}

\subsection{Random walk with drift}
\label{appendix:forecasterrorsRW}

This section derives the distribution of forecast errors. Note that by definition $y_{t+1}-y_t=\Delta y \sim \mathcal{N}(\mu,K^2)$. To obtain  $\hat \mu$ we assume $m$ sequential independent observations of $\Delta y$, and compute the average. The sampling distribution
of the mean of a normal variable is
\begin{equation}
\hat{\mu} \sim \mathcal{N}(\mu,K^2/m).
\label{eq:distmuhat}
\end{equation}
Moreover, $n_t\sim \mathcal{N}(0,K^{2})$ implies
\begin{equation}
\sum_{i=t+1}^{t+\tau} n_{i}\sim \mathcal{N}(0,\tau K^2).
\label{eq:distNttau}
\end{equation}
Using Eqs. (\ref{eq:distmuhat}) and (\ref{eq:distNttau}) in Eq. (\ref{eq:error}) we see that the distribution of forecast errors is Gaussian
\begin{equation}
\mathcal{E}=\tau(\mu-\hat{\mu})+\sum_{i=t+1}^{t+\tau} n_{i} \sim \mathcal{N}(0, K^2A),
\label{eq:distrE}
\end{equation}
where $A=\tau+\tau^2/m$ \eqref{eq:A}. Eq. \ref{eq:distrE}
implies
\begin{equation}
\frac{1}{\sqrt{A}}\frac{\mathcal{E}}{K} \sim \mathcal{N}(0,1).
\label{eq:QEKisN01}
\end{equation}
Eq. (\ref{eq:distrE}) leads to $E[\mathcal{E}^2]=K^2(\tau+\tau^2/m)$, which appears in more general form in \citet{sampson1991effect}. However we also have to account for the fact that we have to estimate the variance. Since $\hat{K}^2$ is the sample variance of a normally distributed random variable, the following standard result holds 
\begin{equation}
\frac{(m-1) \hat{K}^2}{K^2} \sim \chi^2(m-1).
\label{eq:Khatdist}
\end{equation}
If  $Z \sim \mathcal{N}(0,1)$, $U \sim \chi^2(r)$, and $Z$ and $U$ are independent, then $Z/\sqrt{U/r} \sim t(r)$. Taking $Z$ from Eq. (\ref{eq:QEKisN01}), $U$ from Eq. (\ref{eq:Khatdist})
and assuming independence, we find that
the rescaled normalized forecast errors have a Student $t$ distribution
\begin{equation}
\frac{1}{\sqrt{A}} \frac{\mathcal{E}}{\hat{K}} \sim t(m-1).
\end{equation}
Note that the $t$ distribution has mean 0 but variance $df/(df-2)$, where $df$ are the degrees of freedom. Hence the expected squared rescaled normalized forecast error is
\[
E\left [ \left(\frac{1}{\sqrt{A}}\frac{\mathcal{E}}{\hat{K}} \right)^2\right]=0+Var\left [ \frac{1}{\sqrt{A}}\frac{\mathcal{E}}{\hat{K}}\right]=\frac{m-1}{m-3},
\]
leading to Eq.~(\ref{MSFEmain}) in the main text.

\subsection{Integrated Moving Average}
\label{appendix:forecasterrorsIMA}

Here we derive the distribution of forecast errors given that the true process is an IMA(1,1) with known $\theta$, $\mu$ and $K$ are estimated assuming that the process is a random walk with drift, and the forecasts are made as if the process was a random walk with drift. First note that, from Eq. (\ref{eq:IMA11}),
\[
y_{t+\tau}=y_t+\mu \tau+\sum_{i=t+1}^{t+\tau}[v_i+\theta v_{i-1}].
\]
Using Eq. (\ref{eq:predictionMRW}) to make the prediction implies that
\[
\mathcal{E}=y_{t+\tau}-\hat{y}_{t+\tau}=\tau(\mu-\hat{\mu})+\sum_{i=t+1}^{t+\tau}[v_i+\theta v_{i-1}].
\]
Now we can substitute \[
\hat \mu=\frac{1}{m}\sum_{i=t-m}^{t-1}(y_{i+1}-y_i)=\mu+\frac{1}{m}\sum_{i=t-m}^{t-1}[v_{i+1}+\theta v_{i}]
\]
to obtain
\[
\mathcal{E}=\frac{\tau}{m}\left(- \sum_{i=t-m}^{t-1}[v_{i+1}+\theta v_{i}] \right)+\sum_{i=t+1}^{t+\tau}[v_i+\theta v_{i-1}].
\]
Expanding the two sums, this can be rewritten
\[
\begin{split}
\mathcal{E}=&-\frac{\tau \theta}{m}v_{t-m}-\frac{\tau(1+\theta)}{m} \sum_{i=t-m+1}^{t-1}v_i\\
&+\left(\theta-\frac{\tau}{m}\right) v_{t}+(1+\theta) \sum_{i=t+1}^{t+\tau-1}v_i +v_{t+\tau}. 
\end{split}
\]

Note that the term $v_{t}$ enters in the forecast error both because it has an effect on parameter estimation \emph{and} because of its effect on future noise. Now that we have separated the terms we are left with a sum of independent normal random variables. Hence we can obtain $\mathcal{E} \sim\mathcal{N}(0,\sigma^2 A^*)$, where
\[
\begin{split}
A^*\equiv& \left(\frac{\tau \theta}{m}\right)^{2} +(m-1)\left(\frac{\tau (1+\theta)}{m}\right)^{2}\\
&+\left(\theta-\frac{\tau}{m}\right)^2+(\tau-1)(1+\theta)^2+1.
\end{split}
\]
can be simplified as \eqref{eq:Astar} in the main text.

\begin{figure}[H]
\includegraphics[height=70mm]{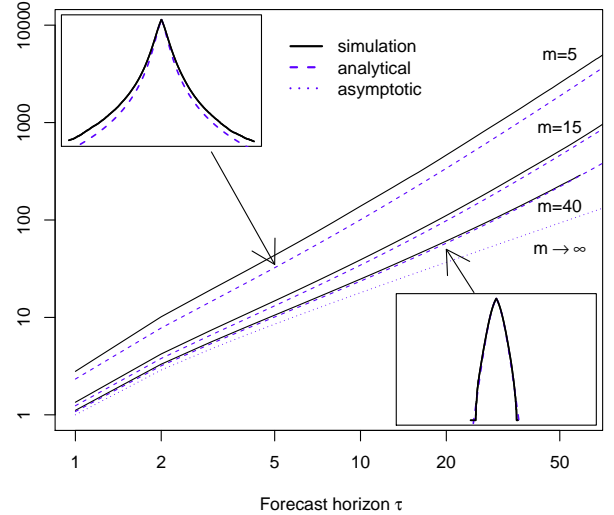}
\caption{\emph{Error growth for large simulations of a IMA(1,1) process}, to check Eq.~(\ref{MSFEmainIMA}) and (\ref{eq:distrtMA}). Simulations are done using 5000 time series of 100 periods, all with with $\mu=0.04$, $K=0.05$, $\theta=0.6$. The insets show the distribution of forecast errors, as in Fig.~\ref{fig:tailplotcompare}, for $m=5,40$}
\label{fig:robustIMA}
\end{figure}

To obtain the results with estimated (instead of true) variance (Eq.~(\ref{MSFEmainIMA}) and (\ref{eq:distrtMA})), we follow the same procedure as in Appendix \ref{appendix:forecasterrorsRW}, which assumes independence between the error and the estimated variance.  Fig.~\ref{fig:robustIMA} shows that the result is not exact but works reasonably well if $m>15$.

\section{Robustness checks}

\subsection{Size of the learning window}
\label{section:robustm}

\begin{figure}[H]
\includegraphics[height=70mm]{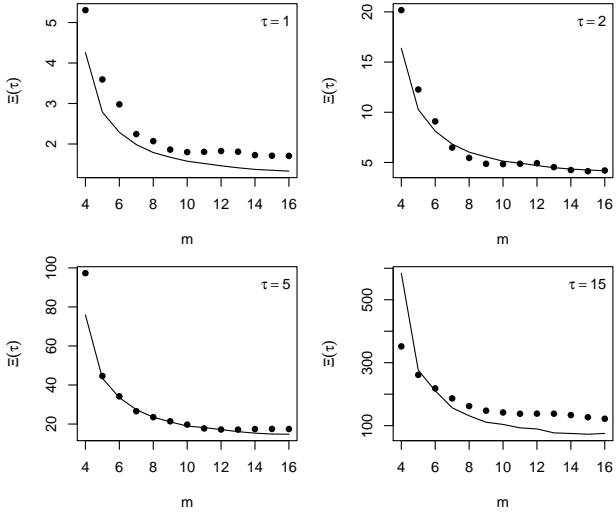}
\caption{\emph{Empirical mean squared normalized forecast errors as a function of the size of learning window for different forecast horizons}. The dots are the empirical errors and the plain lines are those expected if the true model was an IMA(1,1) with $\theta_m =0.63$.}
\label{fig:robustm2}
\end{figure}

As a check on the results presented in Section~\ref{sec:m} we test the dependence of the forecast errors on the sample window $m$ for several different forecast horizons.  The results are robust to a change of the size of learning window $m$. It is not possible to go below $m=4$ because when $m=3$ the Student distribution has $m-1=2$ degrees of freedom, hence an infinite variance. Note that to make forecasts using a large $m$ only the datasets which are long enough can be included.  The results for a few values of $m$ are shown in Fig.~\ref{fig:errorgrowthrobustW}. Fig.~\ref{fig:robustm2}  shows that the normalized mean squared forecast error consistently decreases as the learning window increases.

\subsection{Data selection}

\begin{figure}[H]
\includegraphics[height=70mm]{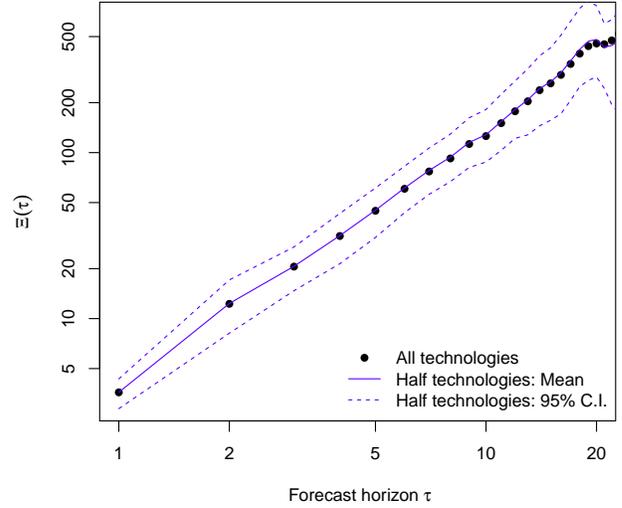}
\caption{\emph{Robustness to dataset selection}. Mean squared normalized forecast errors as a function of $\tau$ when using only half of the technologies (26 out 53), chosen at random. The 95\% confidence intervals, shown as dashed lines, are for the mean squared normalized forecast errors when we randomly select 26 technologies.}
\label{fig:robustNdataset}
\end{figure}

We have checked how the results change when about half of the technologies are randomly selected and removed from the dataset. The shape of the normalized mean squared forecast error growth does not change and is shown in Fig.~\ref{fig:robustNdataset}. The procedure is based on 10000 random trials selecting half the technologies.

\subsection{Increasing $\tau_{max}$}
\label{section:appendixtaumax}

\begin{figure}[H]
\includegraphics[height=70mm]{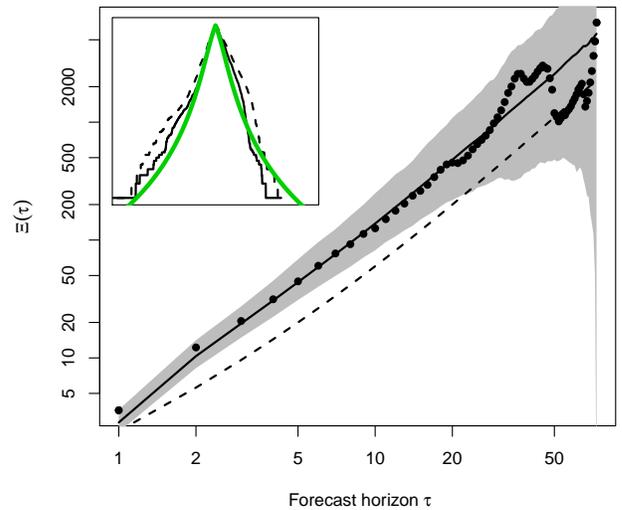}
\caption{\emph{Robustness to increasing $\tau_{max}$}. Main results (i.e as in Fig.~\ref{fig:errorgrowth} and \ref{fig:tailplotcompare}) using $\tau_{max}=73$. We use $\theta=0$ and $\theta=0.63$.}
\label{fig:errorgrowthtau72}
\end{figure}

In the main text we have shown the results for a forecast horizon up to $\tau_{max}=20$. Moreover, we have used only the forecast errors up to $\tau_{max}$ to construct the empirical distribution of forecast errors in Fig.~\ref{fig:tailplotcompare} and to estimate $\theta$ in Appendix \ref{section:thetaselection}. Fig.~\ref{fig:errorgrowthtau72} shows that if we use all the forecast errors up to the maximum with $\tau = 73$ the results do not change significantly.

\subsection{Heavy tail innovations}
\label{section:heavytailinnovations}

\begin{figure}[H]
\includegraphics[height=70mm]{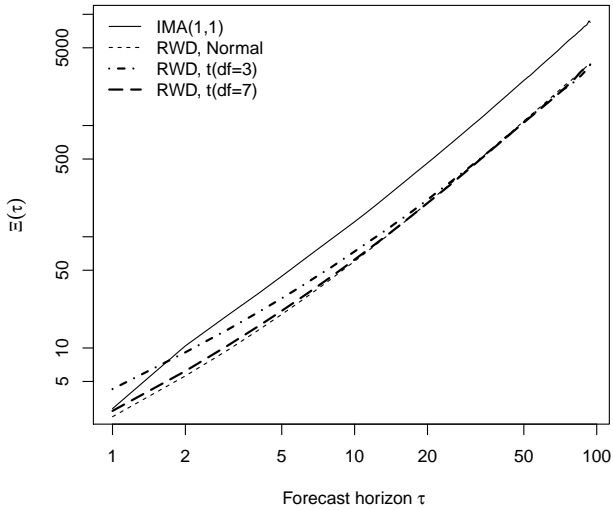}
\caption{\emph{Effect of fat tail innovations on error growth.} The figure shows the growth of the mean squared normalized forecast errors for four models, showing that introducing fat tail innovations in a random walk with drift (RWD) mostly increases errors only at short horizons.}
\label{fig:robustStud}
\end{figure}

To check the effect of non-normal noise increments on $\Xi(\tau)$ we simulated random walks with drift with noise increments drawn from a Student distribution with 3 or 7 degrees of freedom. Fig.~\ref{fig:robustStud} shows that fat tail noise increments do not change the long horizon errors very much.  While the IMA(1,1) model produces a parallel shift of the errors at medium to long horizons, the Student noise increments generate larger errors mostly at short horizons. Thus fat-tail innovations are not the most important source of discrepancy between the geometric random walk model and the empirical data.

\section{Procedure for selecting the autocorrelation parameter $\theta$}
\label{section:thetaselection}

\begin{figure}[h]
\centering
\includegraphics[height=35mm]{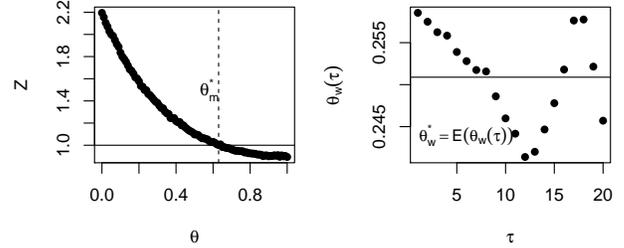}
\caption{\emph{Estimation of $\theta$ as a global parameter}}
\label{fig:thetachoice1}
\end{figure}

\begin{figure}[h]
\centering
\includegraphics[height=35mm]{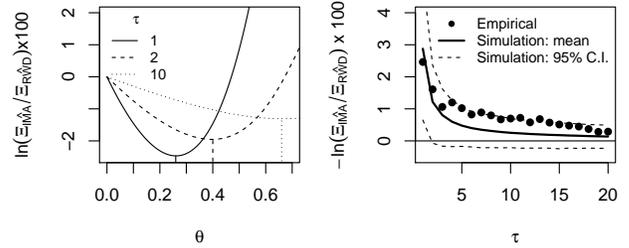}
\caption{\emph{Using the IMA model to make better forecasts}. The right panel uses $\theta=0.25$}
\label{fig:thetachoice2}
\end{figure}

We select $\theta$ in several ways.  The first method is to compute a variety of weighted means for the $\tilde \theta_j$ estimated on individual series. The main problem with this approach is that for some technology series the estimated $\theta$ was very close to 1 or -1, indicating mis-specification or estimation problems.  After removing these 8 technologies the mean with equal weights for each technology is $0.27$ with standard deviation $0.35$.  We can also compute the weighted mean at each forecast horizon, with the weights being equal to the share of each technology in the number of forecast errors available at a given forecast horizon. In this case the weighted mean $\theta_w(\tau)$ will not necessarily be constant over time.  Fig.~\ref{fig:thetachoice1} (right) shows that $\theta_w(\tau)$ oscillates between 0.24 and 0.26.   Taking the average over the first 20 periods gives $\theta_w=\frac{1}{20}\sum_{\tau=1}^{20}\theta_w(\tau)=0.25$.
When doing this we do not mean to imply that our formulas are valid for a system with heterogenous $\theta_j$; we simply propose a best guess for a universal $\theta$.

The second approach is to select $\theta$ in order to match the errors.  As before we generate many artificial data sets using the IMA(1,1) model.   Larger values of $\theta$ imply that using the simple random walk model to make the forecasts will result in higher forecast errors. Denote by $\Xi(\tau)_{empi}$ the empirical mean squared normalized forecast error as depicted in Fig.~\ref{fig:errorgrowth}, and by $\Xi(\tau)_{sim,\theta}$ the expected mean squared normalized forecast error obtained by simulating IMA(1,1) datasets 3,000 times with a particular global value of $\theta$ and taking the average. We study the ratio of these two, averaged over all $1 \dots\tau_{max}=20$ periods, i.e. $Z(\theta) = \frac{1}{20}\sum_{\tau=1}^{20} \frac{\Xi(\tau)_{empi}}{\Xi(\tau)_{sim,\theta}}.$
The values are shown in Fig.~\ref{fig:thetachoice1} (left).  
The value at which $|Z-1|$ is minimum is at $\theta_{m}=0.63$.

We also tried to make forecasts using the IMA model to check that forecasts are improved: which value of $\theta$ allows the IMA model to produce better forecasts? We apply the IMA(1,1) model with different values of $\theta$ to make forecasts (with the usual estimate of the drift term $\hat{\mu}$) and study the normalized error as a function of $\theta$.   We record the mean squared normalized error and repeat this exercise for a range of values of $\theta$. The results for horizons 1,2, and 10 are reported in Fig.~\ref{fig:thetachoice2} (left).  This shows that the best value of $\theta$ depends on the time horizon $\tau$. The curve shows the mean squared normalized forecast error at a given forecast horizon as a function of the value of $\theta$ assumed to make the forecasts. The vertical lines show the minima at 0.26, 0.40, and 0.66.   Given that the mean squared normalized forecast error increases with $\tau$, to make the curves fit on the plot the values are normalized by the mean squared normalized forecast error using $\theta=0$. We also see that as the forecast horizon increases the improvement from taking the autocorrelation into account decreases (Fig. \ref{fig:thetachoice2}, right), as expected theoretically from an IMA process.  Note that the improvement in forecasting error is only a few percent, even for $\tau = 1$, indicating that this makes little difference.

\section{Comparison of the empirical distribution of rescaled errors to the predicted Student distribution}  
\label{sec:statTest}

In this section we check whether the deviations of the empirical forecast errors from the predicted theoretical distribution shown in Fig.~\ref{fig:tailplotcompare} are consistent with statistical sampling error.   For a given value of $\theta$ we generate a surrogate data set and surrogate forecasts mimicking our empirical data as described at the end of Section~\ref{section:hindcasting}.  We then construct a sample surrogate (cumulative) distribution $P_k$ for the pooled rescaled errors $\epsilon^*$ of Eq.~(\ref{eq:distrtMA}).  We measure the distribution $P_k$ over 1,000 equally spaced values $x_k$ on the interval $[-15;15]$.   $P_k$ is estimated by simply counting the number of observations less than $x_k$.  This is then compared to the predicted Student distribution $t_k$ by computing the difference $\Delta_k  = P_k - t_k$ between the surrogate distribution and the Student distribution in each interval.  We measure the overall deviation between the surrogate and the Student using three different measures of deviation: $\sum_k |\Delta_k|$, $\sum_k (\Delta_k)^2$, and $\max \Delta_k$.  We then repeat this process 10,000 times to generate a histogram for each of the measures above, and compare this to the measured value of the deviation for the real data.

\begin{figure}[H]
\includegraphics[height=55mm]{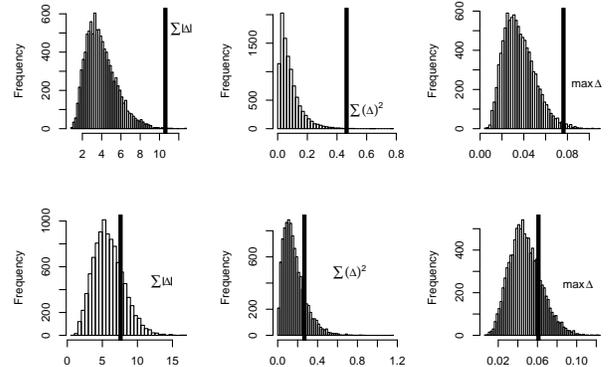}
\caption{\emph{Expected deviations of the distribution of the rescaled variable $\epsilon^*$ of Eq.~(\ref{eq:distrtMA}) from the Student distribution for hindcasting experiments as we do here using a dataset with the same properties as ours}. The histograms show the sampling distribution of a given statistic and the thick black line shows the empirical value on real data. The simulations use $\theta=0.25$ (3 upper panels) and $\theta=0.63$ (3 lower panels).}
\label{fig:Z2}
\end{figure}

Results for doing this for $\theta_w=0.25$ and $\theta_{m}=0.63$ are reported in Fig.~\ref{fig:Z2}.  For $\theta_w$ the resulting $p$-values (the shares of random datasets with a deviation higher than the empirical deviation) are $(0.001, 0.002, 0.011)$ respectively using $(\sum_k |\Delta_k|$, $\sum_k (\Delta_k)^2$, $\max \Delta_k)$ to measure the deviation. In contrast for $\theta_m =0.63$ the $p$-values are $(0.21, 0.16, 0.20)$.  Thus $\theta_m = 0.63$ is accepted and $\theta_w = 0.25$ is rejected.  The uncorrelated case $\theta = 0$ is rejected even more strongly.

\section{A trend extrapolation of solar energy capacity}
\label{solarAppendix}

In this paper we have been concerned with forecasting costs.  For some applications it is also useful to forecast production.  Our exploratory work so far suggests that, while the same basic methods can be applied, production seems more likely to deviate systematically from increasing exponentially. Nonetheless, \cite{nagy2013statistical} found that as a rough approximation most of the technologies in our data set can be crudely (but usefully) approximated as having exponentially increasing production for a long span of their development cycle, and solar PV is no exception.  Trend extrapolation can add perspective, even if it comes without good error estimates, and the example we present below motivates the need for more work to formulate better methods for assessing the reliability of production forecasts (for an example, see \citet{shlyakhter1994quantifying}).

Many analysts have expressed concerns about the time required to build the needed capacity for solar energy to play a role in reducing greenhouse gas emissions.  The "hi-Ren" (high renewable) scenario of the International Energy Agency assumes that PV will generate $16\%$ of total electricity\footnote{
Electricity generation uses about 40\% of the world's primary energy but is expected to grow significantly.}
 in 2050; this was recently increased from the previous estimate of only $11\%$.  As a point of comparison, what do past trends suggest?
 
\begin{figure}[H]
\includegraphics[height=70mm]{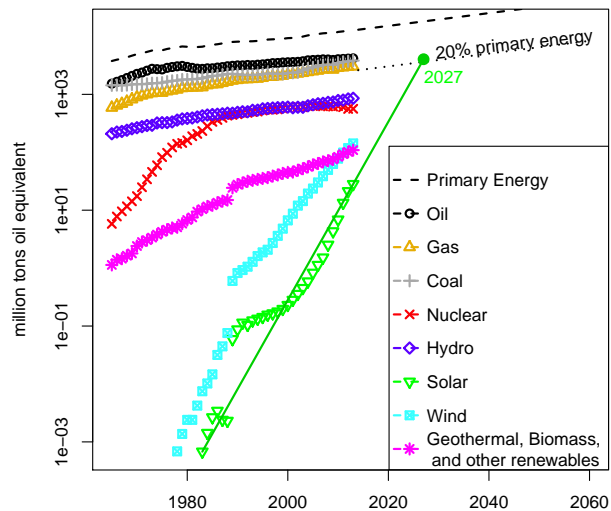}
\caption{\emph{Global energy consumption due to each of the major sources} from BP Statistical Review of World Energy \citep{BPreview}. Under a projection for solar energy obtained by fitting to the historical data the target of $20\%$ of global  primary energy is achieved in 2027.}
\label{energyUsageComparison}
\end{figure}

Though estimates vary, over the last ten years cumulative installed capacity of PV has grown at an impressive rate.  According to BP's Statistical Review of World Energy 2014, during the period from 1983-2013 solar energy as a whole grew at an annual rate of 42.5\%  and in 2014 represented about 0.22\% of total primary energy consumption, as shown in Fig.~\ref{energyUsageComparison}.  
By comparison total primary energy consumption grew at an annual rate of
2.6\% over the period 1965-2013.  Given that solar energy is an intermittent source, it is much easier for it to contribute when it supplies only a minority of energy:  new supporting technologies will be required once it becomes a major player.   If we somewhat arbitrarily pick $20\%$ as a target, assuming both these trends continue unaltered, a simple calculation shows that this would be achieved in about 13.7 years\footnote{
In this deterministic setting, the time to meet this goal is the solution for $t$ of $0.0022 (1.425)^{t} = 0.2 (1.026)^t$.}.
That is, under these assumptions in 2027 solar would represent $20\%$ of energy consumption. Of course this is only an extrapolation, but it puts into perspective claims that solar energy cannot play an essential role in mitigating global warming on a relatively short timescale.

Of course the usual caveats apply, and the limitations of such forecasting is evident in the historical series of Fig.~\ref{energyUsageComparison}.  The increase of solar is far from smooth, wind has a rather dramatic break in its slope in roughly 1988, and a forecast for nuclear power made in 1980 based on production alone would have been far more optimistic than one today.  It would be interesting to use a richer economic model to forecast cost and production simultaneously, but this is beyond the scope of this paper.  The point here was simply to show that if growth trends continue as they have in the past significant contributions by solar are achievable.

\bibliographystyle{agsm}
\bibliography{bib-PerfCurves}

\end{document}